\documentclass[twocolumn]{aastex63}

\usepackage{graphicx}
\usepackage{ulem}
\usepackage{amsmath}
\usepackage{wrapfig}
\usepackage[thinlines]{easytable}
\usepackage{tikz}
\usepackage{hyperref}

\newcommand{\be}{\begin{eqnarray}}
\newcommand{\ee}{\end{eqnarray}}
\newcommand{\lp}{\left(}
\newcommand{\rp}{\right)}
\newcommand{\lb}{\left[}
\newcommand{\rb}{\right]}

\begin{document}
\shorttitle{}
\shortauthors{Piro \& Mockler}

\title{Late-time Evolution and Instabilities of Tidal Disruption Disks}

\correspondingauthor{Anthony L.\ Piro}
\email{piro@carnegiescience.edu}

\author[0000-0001-6806-0673]{Anthony L.\ Piro}
\affiliation{The Observatories of the Carnegie Institution for Science, Pasadena, CA 91101, USA}

\author[0000-0001-6350-8168]{Brenna Mockler}
\affiliation{The Observatories of the Carnegie Institution for Science, Pasadena, CA 91101, USA}

\begin{abstract}
Observations of tidal disruption events (TDEs) on a timescale of years after the main flare show evidence of continued activity in the form of optical/UV emission, quasi-periodic eruptions, and delayed radio flares. Motivated by this, we explore the time evolution of these disks using semi-analytic models to follow the changing disk properties and feeding rate to the central black hole (BH). We find that thermal instabilities typically begin $\sim100\,{\rm days}$ after the TDE, causing the disk to cycle between high and low accretion states for up to $\sim10\,{\rm yrs}$. {The high state is super-Eddington, which may be associated with outflows that eject $\sim10^{-3}-10^{-1}\,M_\odot$ over $\sim1-2\,{\rm days}$ with a range of velocities of $\sim0.03-0.3c$. Collision between these mass ejections may cause radio flares. In the low state, the accretion rate slowly grows over months to years as continued fallback accretion builds the disk's mass. In this phase, the disk has a luminosity of $\sim10^{41}-10^{42}\,{\rm erg\,s^{-1}}$ in the optical/UV as seen in some late-time observations. Although the accretion cycles we find occur for a typical $\alpha$-disk, in nature the disk could be stabilized by other effects such as the disk's magnetic field or heating from fallback accretion, the latter of which we explore. Thus higher cadence optical/UV observations along with joint radio monitoring will be key for following the disk state and testing these models.}
\end{abstract}


\section{Introduction}
\label{sec:intro}

Tidal disruption events (TDEs) occur when an unfortunate star wanders too close to a supermassive black hole (BH), producing a spectacular electromagnetic transient  \citep{Rees88,Phinney89}. For a TDE to be observable, the tidal disruption radius $R_t$ must be exterior to the BH's gravitational radius \citep[e.g.,][]{MacLeod2012}. Otherwise the star will be swallowed by the BH before it has a chance to be ripped apart. This puts a limit on the TDE-producing BH mass of $M_{\rm BH}\lesssim10^8\,M_\odot$, which means TDEs are well suited for probing BHs and their environments in a mass range that is not usually observed in other supermassive BH studies (e.g., active galactic nuclei). Initial samples of TDEs are already providing mass estimates for BHs in the range of $10^6-10^8\,M_\odot$ \citep[e.g.,][]{Mockler2019}, and soon with the Vera Rubin Observatory the ability to study BH demographics in this way is going to increase exponentially \citep{Bricman2020}.

The TDE is seen as a flare that lasts for many weeks to months, and can be observed across a wide range of electromagnetic wavelengths (often for the same event). In recent years it has become clear that TDE BHs can remain active for many years following the initial flare. One way this is seen is from the optical/UV emission that can persist from the TDE location \citep{vanVelzen19,Mummery2020,Nicholl2024}. Another is the presence of X-ray flares called quasi-periodic eruptions \citep[QPEs,][]{Miniutti2019,Giustini2020,Chakraborty2021,Arcodia2021,Arcodia2022,Miniutti2023b,Arcodia2024}. It was expected that there was a connection between TDEs and QPEs due to host galaxy similarities \citep{Wevers2022}, theoretical arguments \citep{Linial2023,Franchini2023}, and persistent X-ray emission \citep{Chakraborty2021,Miniutti2023a,Quintin2023}, and this was confirmed by an optical TDE that years later produced QPEs \citep{Nicholl2024}.

This emission observed years after TDEs is generally thought to require a long lasting disk. This would naturally produce the persistent optical/UV mission, and in the case of QPEs, a disk would be needed for either the star-disk collision models \citep[e.g.,][]{Linial2023,Franchini2023} or in the disk instability models \citep[e.g.,][]{Pan2022,Kaur2023}. \citet{Shen14} considered the viscous evolution of TDE accretion disks on a timescale of up to $\sim10^4\,{\rm yrs}$. They generally found that after a few months to about a year, a thermal instability would transition the disk to a gas-pressure-dominated low accretion state. This might naturally explain the jet shutoff at $\sim500\,{\rm days}$ from the jetted TDE candidate {\it Swift} J1644+57 \citep{Levan2011,Zauderer2013}, but it makes the late time optical/UV emission difficult to reconcile with the low expected accretion rate at these times. Nevertheless, groups have fit the late time optical/UV and constructed disk models for the QPEs either by treating the disk scaleheight as a free parameter \citep[e.g.,][]{Mummery2024} or by using an alternative viscosity prescription \citep[e.g.,][which we discuss in more detail below]{vanVelzen19}. {\citet{Lu2022} updated the models from \citet{Shen14} by implementing a more detailed opacity treatment and a higher viscosity.} \citet{Linial24} also considered the evolution of TDE disks over long time scales, but focused on potential feeding from the ablation of the QPE-producing star.

Many TDEs have also been seen to exhibit radio flares $\sim100-3000\,{\rm days}$ after the main optical/UV emission \citep{Alexander2020,Horesh2021,Horesh2021b,Cendes2022,Goodwin2023,Goodwin2023b,Christy2024}. As surveys have become more extensive, it seems this may be a common feature \citep{Cendes2024,Anumarlapudi2024}. High-speed outflows or jet interactions with the circumnuclear medium at early times could produce radio emission via synchrotron radiation \citep[e.g.,][]{Chevalier1998,Barniol2013,Alexander2016}, but this has difficulty explaining the radio flares that occur with especially long delays of $\gtrsim100\,{\rm days}$. Some possible explanations include misaligned precessing jets \citep{Teboul2023,Lu2024}, a decelerated off-axis jet \citep{Matsumoto2024,Sfaradi2024}, or outflow-cloud interactions \citep{Mou2022,Zhuang2024}. An interesting clue may be that the presence of these flares can be related to changes in accretion state \citep[e.g.,][]{Sfaradi2022}, which again suggests that the late-time evolution of TDE disks is important.

{The {\it r}- and {\it g}-band light curves years after TDEs show variations as well \citep{Mummery2024}. These are higher cadence than the UV data, which makes them more sensitive to detecting these changes, and generally show variations by a factor of $\sim3$. Some events exhibit even larger variations, for example  AT2021mhg and AT2020riz show an increase by a factor of $\sim10$, while AT2018lni shows variations in the {\it g}-band by a factor of $\sim6$. This all points to the late time emission not being as constant as may have been previously believed.}

{Motivated by these issues, we conduct a semi-analytic exploration of TDE accretion disks. In Section~\ref{sec:diskmodel}, we summarize the one-zone model we use in this work. This includes spelling out the guiding equations that are used to solve for the time evolution of the disk. We also explore the impact of a more detailed treatment of the opacity, comparing pure electron scattering to Kramers' and OPAL opacities \citep{OPAL}. We show that our disk models cycle between low and high accretion states and discuss the physics that determines this and whether additional heating sources may stifle this instability. In Section~\ref{sec:comparison}, we compare our calculations to the work of \cite{Shen14}. This is done to confirm the numerical methods we employ (which are discussed in more detail in Appendix~\ref{app:solving}), and also to highlight the changes to the evolution we find due to differences in the treatment of the physics. We show that the higher opacities and viscosities we use are key for finding the accretion cycles that were not highlighted before \citep[similar to the conclusions of][]{Lu2022}.}

{We continue in Section~\ref{sec:disksolutions} by summarizing the variety of disk evolutions we expect as a function of different TDE parameters. This shows that the accretion cycles happen more quickly for lower mass BHs. The high accretion states are super-Eddington and may eject $\sim10^{-3}-10^{-1}\,M_\odot$ on a timescale of $\sim1-2\,{\rm days}$. We compare our results to late time observations Section~\ref{sec:observations}, showing that in the low accretion state our models can match the $\sim10^{42}\,{\rm erg\,s^{-1}}$ luminosities seen in the UV and $\sim10^{41}\,{\rm erg\,s^{-1}}$ luminosities seen in the optical. Furthermore, we expect smaller variations in the optical bands than in higher energy bands, and the factor of $\sim3$ variations we find roughly match observations. We conclude in Section~\ref{Sec:conclusions} with a summary of our results and a discussion of future work.}

\section{One-Zone Disk Model}
\label{sec:diskmodel}

To understand the evolution of an accretion disk left over from a TDE, we use a one-zone disk model \citep[e.g.,][]{Metzger2008,Shen14,Lu2022}. Such a disk is characterized by a total mass $M_d$, characteristic radius $R_d$ (which is roughly the outer radius of the disk), and disk angular momentum
\be
    J_d = (GM_{\rm BH}R_d)^{1/2}M_d,
\ee
where $M_{\rm BH}$ is the BH mass. Such a one-zone model can be a good approximation since the majority of the disk mass and angular momentum are at $R_d$, the location which satisfies $t_\nu\sim t$, where $t_\nu$ is the viscous timescale. Since $t_\nu\sim r^{3/2}$, interior to $R_d$ the disk can be assumed to be in steady state. {In Section~\ref{Sec:conclusions}, we come back to this assumption and discuss whether a one-dimensional model may be needed in some cases.} We next explain the basic ingredients included to solve for the time-dependent disk properties.

\subsection{Fallback Feeding during a TDE}

When a star of mass $M_*$ and radius $R_*$ travels too close to a supermassive BH, it will be tidally disrupted if its pericenter distance $R_p$ is smaller than the tidal disruption radius
\be
    R_t &=& R_*(M_{\rm BH}/M_*)^{1/3}
    \nonumber
    \\
    &=&7.0\times10^{12}
    M_6^{1/3}
    m_*^{-1/3}
    r_*\,{\rm cm},
\ee
where $M_{\rm BH}$ is the mass of the BH, $M_6=M_{\rm BH}/10^6\,M_\odot$, $m_*=M_*/M_\odot$, and $r_*=R_*/R_\odot$. The depth of the star's plunge is usually described by the parameter $\beta=R_t/R_p$, so that $1\lesssim\beta\lesssim R_t/R_s$, where $R_s=2GM_{\rm BH}/c^2$ is the Schwarzschild radius (the last ratio can be larger for a Kerr BH). The critical value for a complete disruption $\beta_c$ depends on the central concentration of the state \citep{Guillochon13,LawSmith2020,Ryu2020}, for example, $\beta_c = 1.85$ for an $n= 4/3$ polytrope and $\beta_c = 0.9$ for an $n= 5/3$ polytrope.

\begin{figure}
\includegraphics[width=0.45\textwidth,trim=0.6cm 0.3cm 1.5cm 0.0cm]{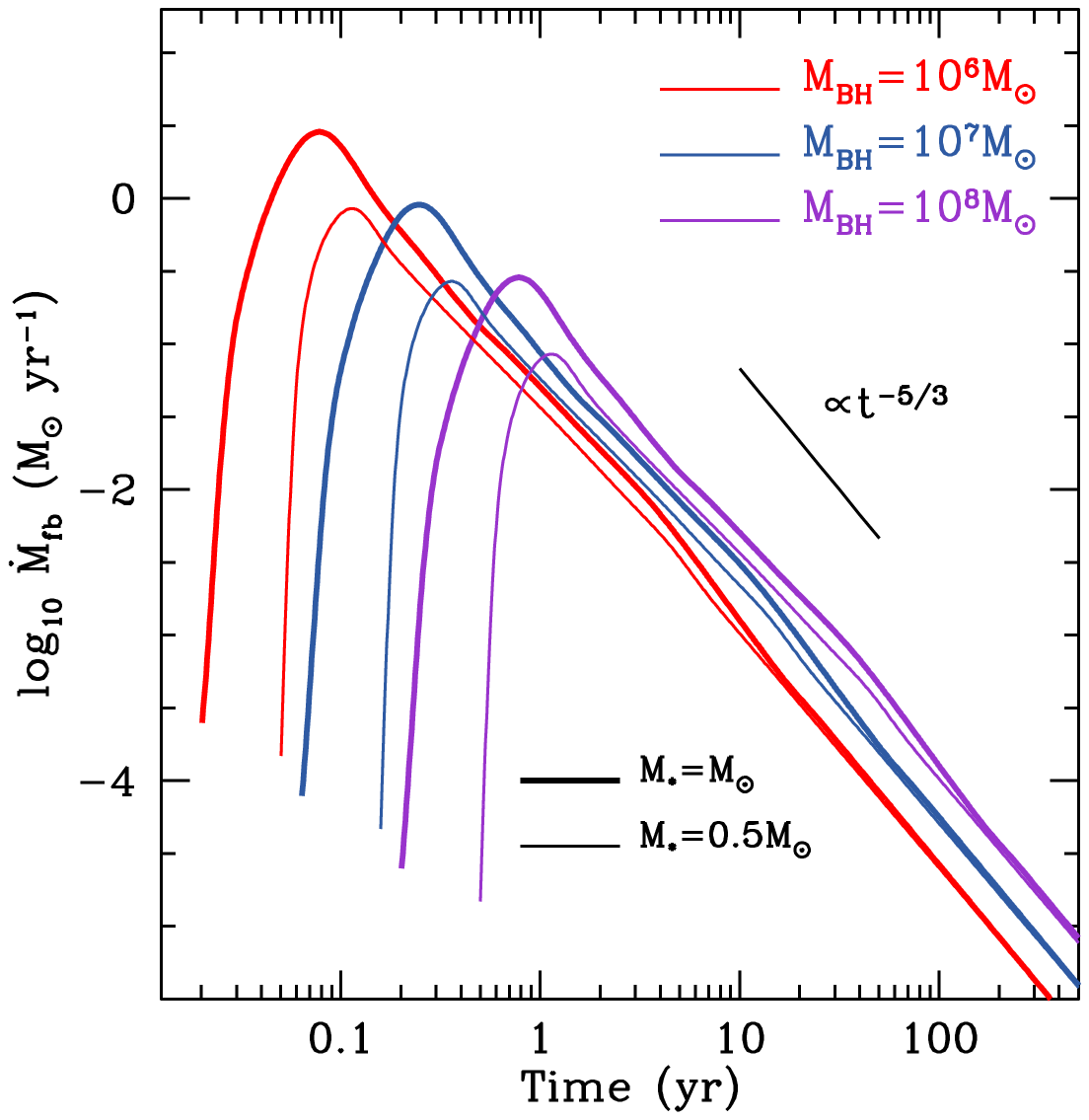}
\caption{Fallback accretion rate for tidal disruptions of stars using the work of \citet{Guillochon13} for $M_*=M_\odot$ with $n=4/3$ and $\beta=1.85$ (thick lines) and $0.5\,M_\odot$ with $n=5/3$ and $\beta=0.9$ (thin lines). The three colors correspond to different values of $M_{\rm BH}$ as indicated.}
\label{fig:fallback}
\end{figure}

The fallback of stellar material can be modeled as a power law for times greater than $t_{\rm fb}$ \citep[e.g.,][]{Rees88,Phinney89},
\be
    \dot{M}_{\rm fb}(t)
        = \frac{M_*}{5t_{\rm fb}}
        \left(\frac{t}{t_{\rm fb}}\right)^{-5/3},
    \label{eq:mdotfb}
\ee
where the fallback timescale is \citep[e.g.][]{Stone13}
\be
    t_{\rm fb}
    &=& \frac{\pi R_t^3}{(2GM_{\rm BH}R_*^3)^{1/2}}
    \nonumber
    \\
    &=& 3.5\times10^6M_6^{1/2}m_*^{-1}r_*^{3/2}\,{\rm s}.
    \label{eq:tfb}
\ee
In this expression we use $R_t$ rather than $R_p$ because for full disruptions with $\beta>\beta_c$ the dependence is generally weaker with $\beta$ than what would be expected from analytic arguments \citep{Stone13,Guillochon13,Gafton2019}.
The peak fallback rate is then
\be
    \frac{M_*}{5t_{\rm fb}}
    = 1.8M_6^{-1/2}m_*^2r_*^{-3/2}
    \,M_\odot\,{\rm yr^{-1}}.
\ee
Although these scaling match numerical results for $t>t_{\rm fb}$, we also need the rising $\dot{M}_{\rm fb}(t)$ for the early phases of the disk evolution. Thus we use the numerical results of \citet{Guillochon13} for our full calculations, which we present for three BH masses and two stellar masses in Figure~\ref{fig:fallback}. In each of these cases, we use $\beta=\beta_c$ since we are interested in full disruptions and the $\dot{M}_{\rm fb}(t)$ does not change too greatly for $\beta\gtrsim \beta_c$. In most of this work we generally use $\beta=\beta_c$ unless we state otherwise.

The fallback material circularizes at a radius
\be
    R_c = 2R_p
    = 1.4\times10^{13}
    \beta^{-1}
    M_6^{1/3}
    m_*^{-1/3}
    r_*\,{\rm cm}.
\ee
Thus as the fallback material is incorporated into the accretion disk, it has specific angular momentum
\be
    j_{\rm fb} &=& (GM_{\rm BH}R_c)^{1/2}
    \nonumber
    \\
    &=& 4.3\times10^{22}
    \beta^{-1/2}
    M_6^{2/3}
    m_*^{-1/6}
    r_*^{1/2}\,{\rm cm^2\,s^{-1}}.
    \label{eq:jfb}
\ee
In detail, the fallback stream may interact at a different radius than $R_c$. For example, if the stream and disk are in a similar plane, then the collision may occur near the outer edge of the disk at $R_d$. Nevertheless, the most important issue is the angular momentum contribution, so we use Equation~(\ref{eq:jfb}) for this work.

\subsection{Time Evolution Equations}

The disk is fed via fallback of material from TDE at a rate $\dot{M}_{\rm fb}$ as described above, which then accretes through the disk at a rate $\dot{M}$. Some fraction of this mass transport may go into an outflow, while the remaining mass accretes all the way down to the BH. {The outflow is expected to be launched within the spherization radius $R_{\rm sph}$, where the local accretion luminosity $GM_{\rm BH}\dot{M}/R_{\rm sph}$ exceeds the the Eddington luminosity $4\pi GM_{\rm BH}c/\kappa$ \citep{Begelman1979},
\be
    R_{\rm sph} = \min \lp R_d,\frac{\kappa\dot{M}}{4\pi c} \rp,
\ee
where we use an electron scattering opacity for $\kappa$. Following previous studies \citep[e.g.,][]{Metzger2008,Yuan2014,Hu2022}, we assume these disk outflows cause the mass inflow rate to decrease within $R_{\rm sph}$ as it approaches the BH \citep[e.g.,][]{Blandford1999}
\be
    \dot{M}(r) = \lp \frac{r}{R_{\rm sph}}\rp^p \dot{M},
\ee
where $0<p<1$ is a parameter that controls the strength of the outflow. The total mass outflow rate is thus
\be
    \dot{M}_{\rm out}
    = \lb 1- \lp \frac{R_i}{R_{\rm sph}}\rp^p\rb \dot{M},
    \label{eq:mdotout}
\ee
where $R_i$ is the inner radius of the disk.} Even with an outflow, the total mass loss of the disk at any time adds up to $\dot{M}$, and thus the differential equation
\be
    \frac{dM_d}{dt}
        = \dot{M}_{\rm fb}
        - \dot{M},
    \label{eq:mass}
\ee
describes the mass evolution. {This expression assumes that the mass feeding rate to the disk tracks the fallback rate. If stream collisions are responsible for circularizing material into a disk to begin with \citep[as opposed to, e.g. nozzle shocks, see][]{steinberg_stream-disk_2024}, Lense-Thirring precession by a rapidly spinning BH can delay this process by causing the stream to miss colliding with itself for many orbits \citep{Dai2013,Guillochon2015,jankovic_spin-induced_2024}. The details of this depend on the general relativistic precession versus the hydrodynamic evolution of the stream thickness. The circularization process still need further exploration, so by simply using $\dot{M}_{\rm fb}$ our results are focused on the period after material has begun to circularize into a disk.}

{We next consider the disk's angular momentum evolution. Fallback adds angular momentum with a specific value of $j_{\rm fb}$, but it can also be removed via the outflows. These two processes are represented in the following differential equation,
\be
    \frac{dJ_d}{dt}
        = j_{\rm fb}\dot{M}_{\rm fb}
        - C(GM_{\rm BH}R_{\rm sph})^{1/2}\dot{M}_{\rm out},
    \label{eq:angularmomentum}
\ee
where the constant $C$ is determined by the torque exerted by the wind on the disk. Assuming that the outflow produces no net torque \citep[e.g.,][]{Stone2001}, then the angular momentum losses are due to the specific angular momentum at each disk radius, resulting in
\be
    C = \frac{2p}{2p+1}
\ee
\citep{Kumar08}. In principle this factor can be higher if large scale magnetic fields help transport additional angular momentum loss, but we do not consider these effects in this work.}
%Note that we use $R_{\rm sph}$ rather than $R_d$ in Equation~(\ref{eq:angularmomentum}) to be general enough to include cases where $R_{\rm sph}\lesssim R_d$, but in practice, for the BH mass range we consider in this work, generally $R_{\rm sph}\approx R_d$ or $R_{\rm sph}\ll R_d$.}

\subsection{Disk Structure}
\label{sec:diskstructure}

The mass loss rate $\dot{M}$ is set using a typical vertically-integrated disk model. We quickly summarize the model here for completeness \citep[mostly following][]{Frank02}.

Hydrostatic balance gives a disk thickness of $H = c_s/\Omega$, where $c_s^2=P/\rho$ is the isothermal sound speed, $P$ and $\rho$ are the midplane pressure and density, respective, and $\Omega = (GM_{\rm BH}/R_d^3)^{1/2}$ is the Keplerian angular speed. For the pressure, we include both ideal gas and radiation components, so that
\be
    P = P_g + P_r
    = \frac{\rho k_{\rm B}T}{\mu m_p}
    + \frac{aT^4}{3},
\ee
where $k_{\rm B}$ is Boltmann's constant, $\mu$ is the mean molecular weight ($0.62$ for solar material), $m_p$ is the proton mass, and $a$ is the radiation constant.

The surface density is $\Sigma = \rho H$, and mass conservation through the disk leads to $\dot{M}=3\pi\nu\Sigma$, where $\nu$ is the viscosity. We parameterize the viscosity using the usual $\alpha$-disk model \citep{Shakura73}
\be
    \nu = \alpha c_s H = \alpha \frac{P}{\Omega\rho},
    \label{eq:nu}
\ee
where $\alpha$ is constant with a typical value of $\alpha=0.1$. It has long been known that simple 1D theory results in disks that are viscously and thermally unstable when the disk is radiation-pressure dominated and cooled radiatively \citep{Lightman74,Shakura76}. This can be alleviated by setting $P$ to $P_g$ in Equation~(\ref{eq:nu}) as in \citet{Sakimoto81}, and in fact, such a prescription is employed by \citet{vanVelzen19} when fitting late-time UV emission from TDEs. Whether these instabilities exist for more realistic magnetohydrodynamic simulations of accretion disks in 3D is still not clear \citep[e.g.,][]{Begelman07,Hirose09,Oda09,Jiang13,Mishra16,Sadowski16} and this may even depend on the details of the radiative transfer and opacity \citep{Jiang16}. Here we focus on the more classical case given by Equation~(\ref{eq:nu}) using the total pressure, and we explore the implications if these instabilities really do occur in nature.

The internal energy in the disk is determined by viscous heating $Q^+ = 9\nu\Sigma\Omega^2/8$, radiative cooling $Q^-_{\rm rad}=acT^4/(3\kappa\Sigma)$, where $\kappa$ is the specific opacity, and advective cooling
\be
    Q^-_{\rm adv} = \frac{\dot{M}}{2\pi R_d^2}c_s^2\xi,
\ee
where $\xi$ is the logarithmic entropy gradient. Since $\xi$ is typically of order unity, we set $\xi=1.5$ for this work \citep[e.g.,][]{Watarai2006}. Local energy balance results in
\be
    \frac{9}{8}\nu\Sigma\Omega^2
    = \frac{acT^4}{3\kappa\Sigma}
    + \frac{\dot{M}}{2\pi R_d^2}c_s^2\xi.
    \label{eq:energy}
\ee
We ignore energy lost to the wind, although see the Appendix of \citet{Shen14} for a discussion of the small correction from this effect. {We also do not include energy loss from vertical convection, which may be particularly important when the disk is super-Eddington \citep{Jiang13}.}

\begin{figure}
\includegraphics[width=0.45\textwidth,trim=0.5cm 0.3cm 1.5cm 0.0cm]{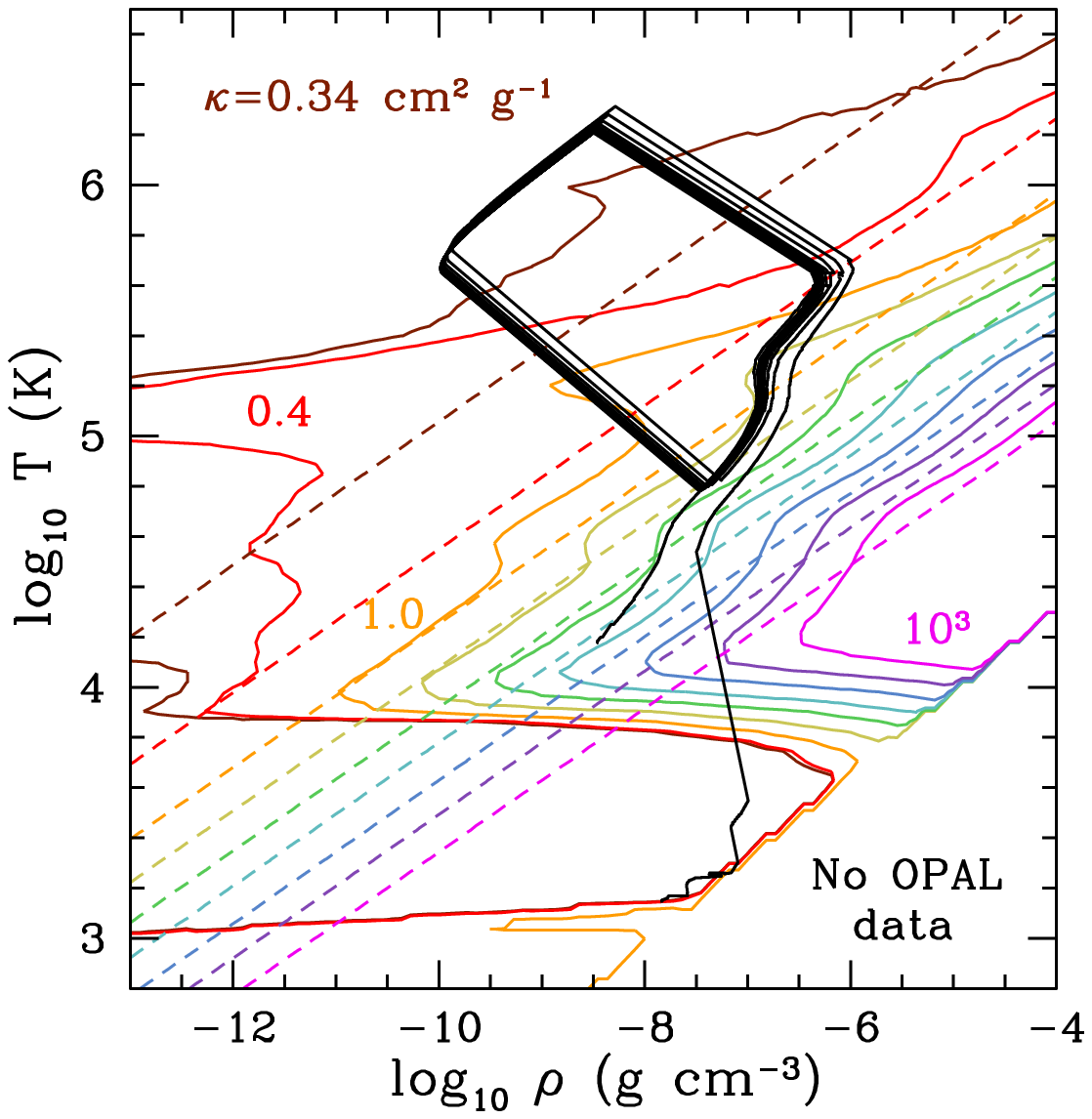}
\caption{Contours of constant opacity for solar-composition material from OPAL (solid colored lines) in comparison to an analytic Kramers plus electron scattering opacity given by Equation~(\ref{eq:opacity}) (dashed colored lines). Dark-red and red curves are contours of constant opacity with values of $0.34\,{\rm cm^2\,g^{-1}}$ and $0.4\,{\rm cm^2\,g^{-1}}$, respectively. The orange through magenta curves are spaced logarithmically in intervals of $10^{0.5}$ from $1.0\,{\rm cm^2\,g^{-1}}$ to $10^3\,{\rm cm^2\,g^{-1}}$. The black solid line shows the trajectory of a fiducial disk solution ($M_{\rm BH}=10^6\,M_\odot$, $M_*=M_\odot$, $\beta=1.85$, and $\alpha=0.1$) evolved over $5,000\,{\rm yrs}$.}
\label{fig:kappa_contour}
\end{figure}

An important issue when solving Equation~(\ref{eq:energy}) is setting $\kappa$ in $Q^-_{\rm rad}$. In previous works addressing similar problems related to long term evolution of TDE disks, this has been set to purely electron scattering $\kappa=\kappa_{\rm es}=0.34\,{\rm cm^2\,g^{-1}}$ \citep[e.g.,][]{Shen14} or a sum of electron scattering and Kramers' opacities \citep[e.g.,][]{Linial24}
\be
    \kappa = \kappa_{\rm es} + \kappa_0\rho T^{-7/2},
    \label{eq:opacity}
\ee
where here $\rho$ and $T$ are assumed to be in cgs units and $\kappa_0=5\times10^{24}\,{\rm cgs}$. In Figure~\ref{fig:kappa_contour}, we plot contours of constant $\kappa$ using Equation~(\ref{eq:opacity}) as dashed colored lines in comparison to the Rosseland mean opacities from OPAL\footnote{https://opalopacity.llnl.gov/} \citep{OPAL} for solar-composition material as solid colored lines. Some important features that are seen for the OPAL opacities that are not captured in a simpler analytic opacity include the following.
\begin{itemize}
    \item An enhanced opacity near $T\approx2\times10^5\,{\rm K}$ due to the iron-opacity ``bump'' \citep[as explored by][]{Jiang16}.
    \item A strong decrease in opacity for $T\lesssim6\times10^3\,{\rm K}$ due to hydrogen recombination.
    \item A slightly enhanced opacity in the range of $\approx~4\times10^4-5\times10^5\,{\rm K}$, which as we will show is especially relevant for the accretion disks we will be considering.
\end{itemize}
Also plotted on Figure~\ref{fig:kappa_contour} as a solid black line is an example disk evolution. We come back to the details of this later (the methods for how this is solved for are described in the following sections), but include it now to highlight the regions of temperature and density space that will be most relevant to this study. In particular, one can directly see the imprint of the the iron bump on the right-side of the trajectory.

\begin{figure}
\includegraphics[width=0.45\textwidth,trim=0.0cm 0.0cm 0.5cm -0.8cm]{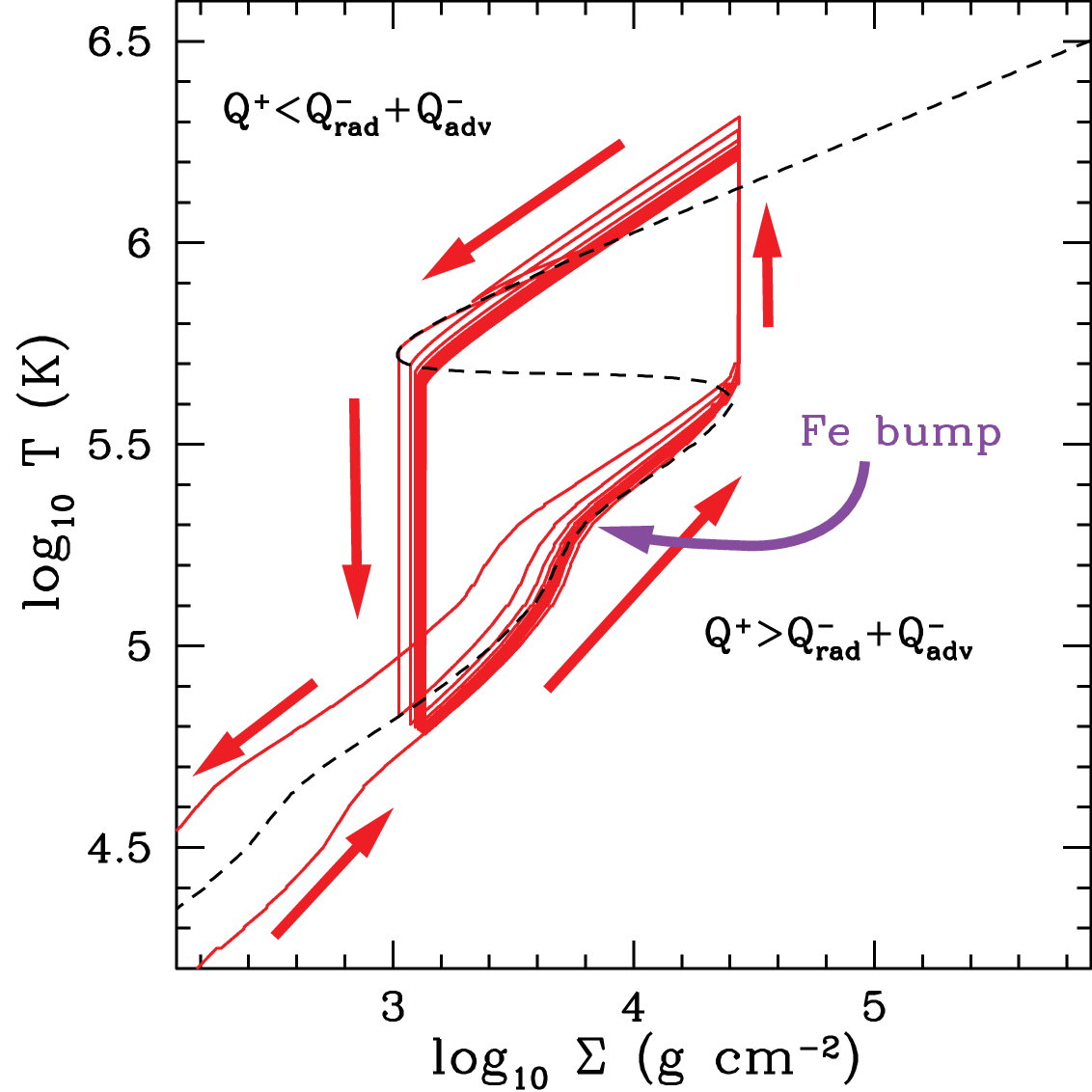}
\caption{The black dashed line represents the surface density $\Sigma$ and temperature $T$ for equilibrium disk solutions using $M_{\rm BH}=10^6\,M_\odot$, $\alpha=0.1$, a fixed radius of $R_d=3\times10^{13}\,{\rm cm}$, and varying $\dot{M}$ from high to low values going from top to bottom. Cooling beats heating on the left of these solutions, and heating beats cooling on the right. The red solid line shows an example disk evolution. The disk starts from the bottom-left corner and evolve toward the right. Once it reaches the instability region, the disk cycles counter clockwise in this space, alternating between low and high states. In purple, we highlight where the iron-opacity bump impacts the evolution.}
\label{fig:sigma_temp_diagram}
\end{figure}

The example disk evolution in Figure~\ref{fig:kappa_contour} circles around in density and temperature due to a thermal instability. Since this plays an important role in our results and future discussions, we focus more on this instability in Figure~\ref{fig:sigma_temp_diagram}. We calculate a series of equilibrium disk solutions by solving Equation~(\ref{eq:energy}) for a range of $\dot{M}$, using $M_{\rm BH}=10^6\,M_\odot$ and $\alpha=0.1$. We fix the disk radius to $R_d=3\times10^{13}\,{\rm cm}$. This results in a locus of solutions shown as a dashed line in Figure~\ref{fig:sigma_temp_diagram}. To the left of the dashed line, cooling exceeds heating, and to the right of the dashed line, heating beats cooling. Even though all of the points on this dashed line are equilibrium solutions, it is well known that regions where $dT/d\Sigma<0$ are thermally unstable \citep{Lightman74,Shakura76}. This is in the regime where the disk is radiation dominated, i.e., $P_{\rm rad}>P_{\rm gas}$, and radiatively cooled, i.e., $Q^-_{\rm rad}>Q^-_{\rm adv}$. In fact, when we extend these models to even lower accretion rates and cooler temperatures ($\lesssim6\times10^3\,{\rm K}$), another ``S-curve'' occurs due to the abrupt change in opacity from hydrogen recombination. This physics is of course important for dwarf nova outbursts \citep{Warner1995}, but would only be exhibited at extremely late times by TDEs and thus outside the scope of this work.

This instability is exhibited by the example time-dependent evolution we show in red (this is the same example model as plotted in Figure~\ref{fig:kappa_contour}). This model starts at low $\Sigma$ and $T$ at the bottom-left corner of the plot and then evolves toward the top right as the disk mass builds. It then goes through cycling behavior when it reaches the unstable region. After circling counterclockwise many times, eventually, once $\dot{M}_{\rm fb}$ is sufficiently low, the disk continues to evolve in the low state toward the bottom-left corner. Note that the black and red lines do no coincide exactly because in the time-dependent model $R_d$ is allowed to vary, while for the equilibrium solutions $R_d$ is fixed. Nevertheless, the equilibrium solution roughly predicts where the instability arises. Finally, we again highlight the impact of the iron-opacity bump (which occurs where the purple arrow points). This essentially creates a two-tiered low state, which is an important property of the time evolution we will come back to later.

\subsection{Fallback Heating}
\label{sec:fallbackheating}

A potentially important physical effect that we do not include in our main models is the interaction of the fallback accretion with the spreading accretion disk. This collision can add additional heating to the disk and modify the energy balance given by Equation~(\ref{eq:energy}). How well the fallback stream is thermalized depends on the ram pressure of the incoming material, the pressure in the disk, and the trajectory of the fallback which is related to the hydrodynamics and dynamics of the stream and disk, so we save a detailed treatment of this for future work. Nevertheless, the heating rate should be proportional to $\dot{M}_{\rm fb}v_{\rm fb}^2/2$, where $v_{\rm fb}\approx(GM_{\rm BH}/R_d)^{1/2}$. {Note that a similar prescription for this heating was used by \citet{Lu2022}.} We use $R_d$ rather than $R_c$ for setting this velocity, which assumes that the incoming stream is in a similar plane to the disk and thus will collide somewhere near the outer radius of the disk. Using this, we estimate the impact of including such effects by rewriting the energy equation as
\be
    \frac{9}{8}\nu \Sigma \Omega^2
    + \frac{\eta\dot{M}_{\rm fb}}{4\pi R_d^2}v_{\rm fb}^2
    = \frac{acT^4}{3\kappa\Sigma}
    + \frac{\dot{M}}{2\pi R_d^2}c_s^2\xi,
    \label{eq:energy_update}
\ee
where $\eta\lesssim1$ is a parameter that sets the efficiency of thermalization. Implicit in this equation is the assumption that the heat will spread fairly quickly around the disk even though the stream collision occurs at a single point.

\begin{figure}
\includegraphics[width=0.45\textwidth,trim=0.5cm 0.3cm 1.5cm 0.0cm]{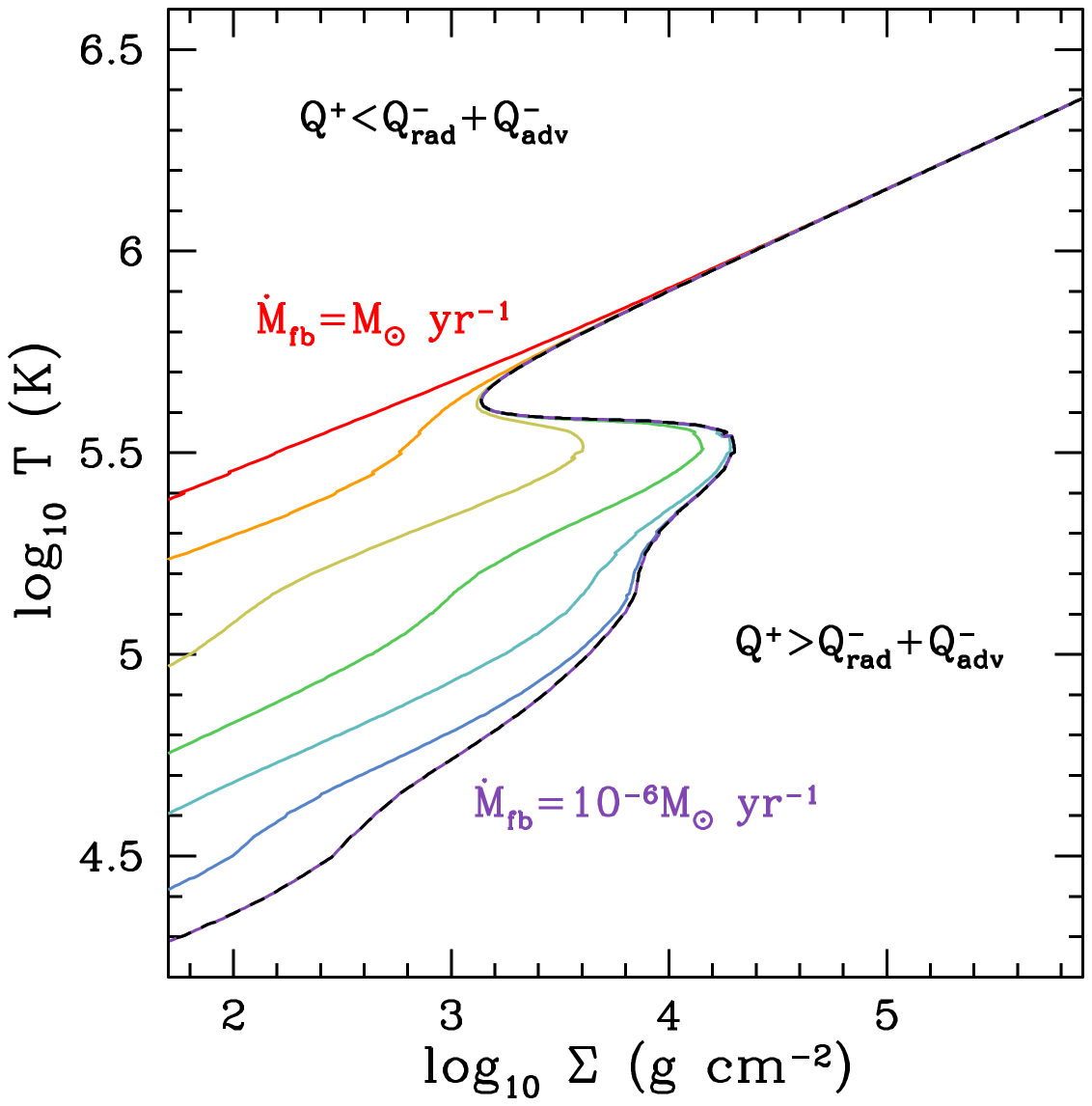}
\caption{The black dashed line matches the same line from Figure~\ref{fig:sigma_temp_diagram}. The solid color lines show the new equilibrium solutions when heating from fallback accretion is included using Equation~(\ref{eq:energy_update}) (all using $\eta=1$, $M_{\rm BH}=10^6\,M_\odot$, $\alpha=0.1$, a fixed radius of $R_d=3\times10^{13}\,{\rm cm}$, and varying $\dot{M}$ from high to low values going from top to bottom). The different colors correspond to varying the fallback accretion by factors of $10$ from $\dot{M}_{\rm fb}=M_\odot\,{\rm yr^{-1}}$ (red line) to $\dot{M}_{\rm fb}=10^{-6}\,M_\odot\,{\rm yr^{-1}}$ (purple line). Note that the black dashed and purple lines are basically coincident.}
\label{fig:sigma_temp_heating}
\end{figure}

In Figure~\ref{fig:sigma_temp_heating}, we recalculate the equilibrium solutions from Figure~\ref{fig:sigma_temp_diagram}, but now including fallback heating using the updated energy Equation~(\ref{eq:energy_update}) with $\eta=1$ and varying the value of $\dot{M}_{\rm fb}$. This shows that when $\dot{M}_{\rm fb}=M_\odot\,{\rm yr^{-1}}$, for this particular example, the instability no longer occurs. As we decrease $\dot{M}_{\rm fb}$, the instability reappears, and at $\dot{M}_{\rm fb}=10^{-6}\,M_\odot\,{\rm yr^{-1}}$, the equilibrium solutions are basically the same as not including fallback heating at all (the black dashed line).

Note that for this treatment $\eta$ and $\dot{M}_{\rm fb}$ are degenerate with one another, so the exact fallback value where the instability disappears depends on $\eta$. We conducted a series of time evolution calculations with different levels of heating (not presented in this work) and found that for high $\eta$ the instability can be removed at early times, which causes $\dot{M}\approx \dot{M}_{\rm fb}$. {This is similar to the results of \citet{Linial24}, who also found that the disk can be stabilized with an additional heating source, but in their case this was from the shock by the QPE-generating star colliding with the disk.}

{\citet{Bonnerot2021} studied stream interactions and found that most of the orbital energy is dissipated by shocks that are away from the disk if there is a strong stream self-crossing shock that diverts the fallback in a somewhat spherical manner. This suggests that $\eta\ll1$, which would make disk stabilization from fallback difficult. This is not a solved problem though, and these uncertainties show that future work is needed to better understand the interactions between the fallback gas and an existing disk so that the amount of heating can be better assessed.}

\section{Comparisons to Previous Work}
\label{sec:comparison}

\begin{figure}
\includegraphics[width=0.45\textwidth,trim=0.5cm 0.3cm 1.5cm 0.5cm]{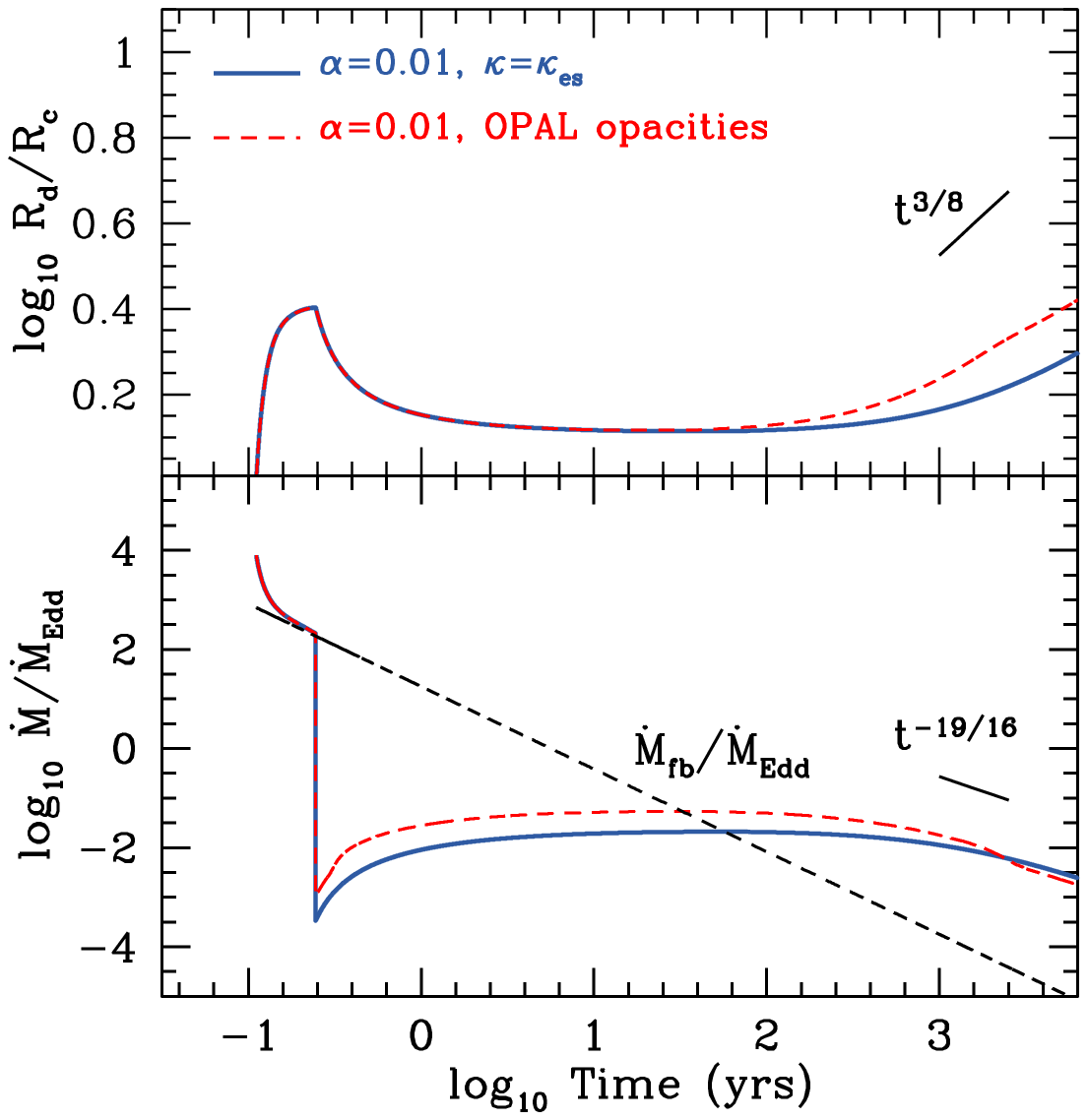}
\includegraphics[width=0.45\textwidth,trim=0.5cm 0.3cm 1.5cm 0.5cm]{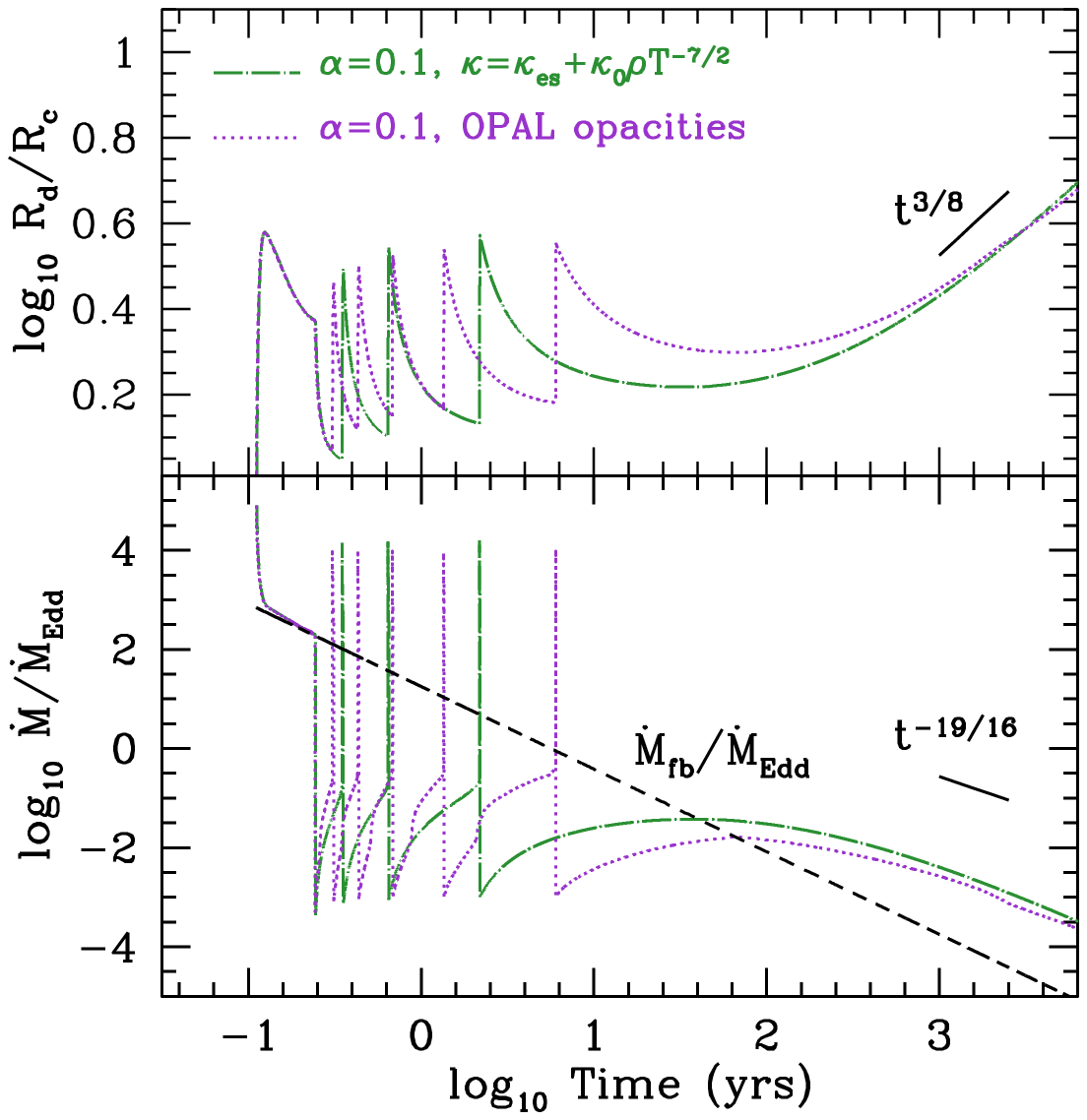}
\caption{Comparison of the disk radius and accretion evolution for four different models. In each case, we use $M_{\rm BH}=10^6\,M_\odot$, $M_*=M_\odot$, and $\beta=1$ with the analytic fallback rate given in Equation~(\ref{eq:mdotfb}). In the upper plot, we set $\alpha=0.01$ and compare $\kappa=\kappa_{\rm es}$ \citep[blue solid lines, meant to mimick the fiducial model from][]{Shen14} with $\kappa$ set by OPAL opacities (red dashed lines). In the lower plot, we set $\alpha=0.1$ and compare $\kappa$ using an analytic opacity given by Equation~(\ref{eq:opacity}) (green dot-dashed lines) and $\alpha=0.1$ with $\kappa$ again set by OPAL opacities (purple dotted lines). The late time power-law behavior was derived in \citet{Shen14}. The dashed black line indicates the fallback accretion rate that is feeding the disk.}
\label{fig:comparison}
\end{figure}

To find the disk evolution, we solve the differential Equations~(\ref{eq:mass}) and (\ref{eq:angularmomentum}) explicitly, where $\dot{M}$ is set using energy balance as given by Equation~(\ref{eq:energy}) and described in Section~\ref{sec:diskstructure}. In Appendix~\ref{app:solving}, we describe the numerical scheme used to solve for the disk structure and time evolution. To check our approach, in this section we make comparisons in some simplified limits.

To facilitate comparisons to \citet{Shen14}, we had to update some of the prescriptions we use in our disk model (our work is set to be consistent with \citealp{Frank02}). To quickly summarize the differences, these are $\Sigma=2\rho H$, $\nu = 2\alpha c_s H/3$, $Q^+ = 9\nu\Sigma\Omega^2/4$, $Q^-_{\rm rad}=4acT^4/(3\kappa\Sigma)$, and setting $\xi=1$. {We also assume here that all accretion above the Eddington rate is ejected rather than use the radially varying mass loss rate as described by Equation~(\ref{eq:mdotout}).} We use these values and relations for the remainder of this section, but for all other calculations shown in this work we use the prescriptions summarized in Section~\ref{sec:diskmodel}.

\begin{figure}
\includegraphics[width=0.45\textwidth,trim=0.5cm 0.3cm 1.5cm 0.0cm]{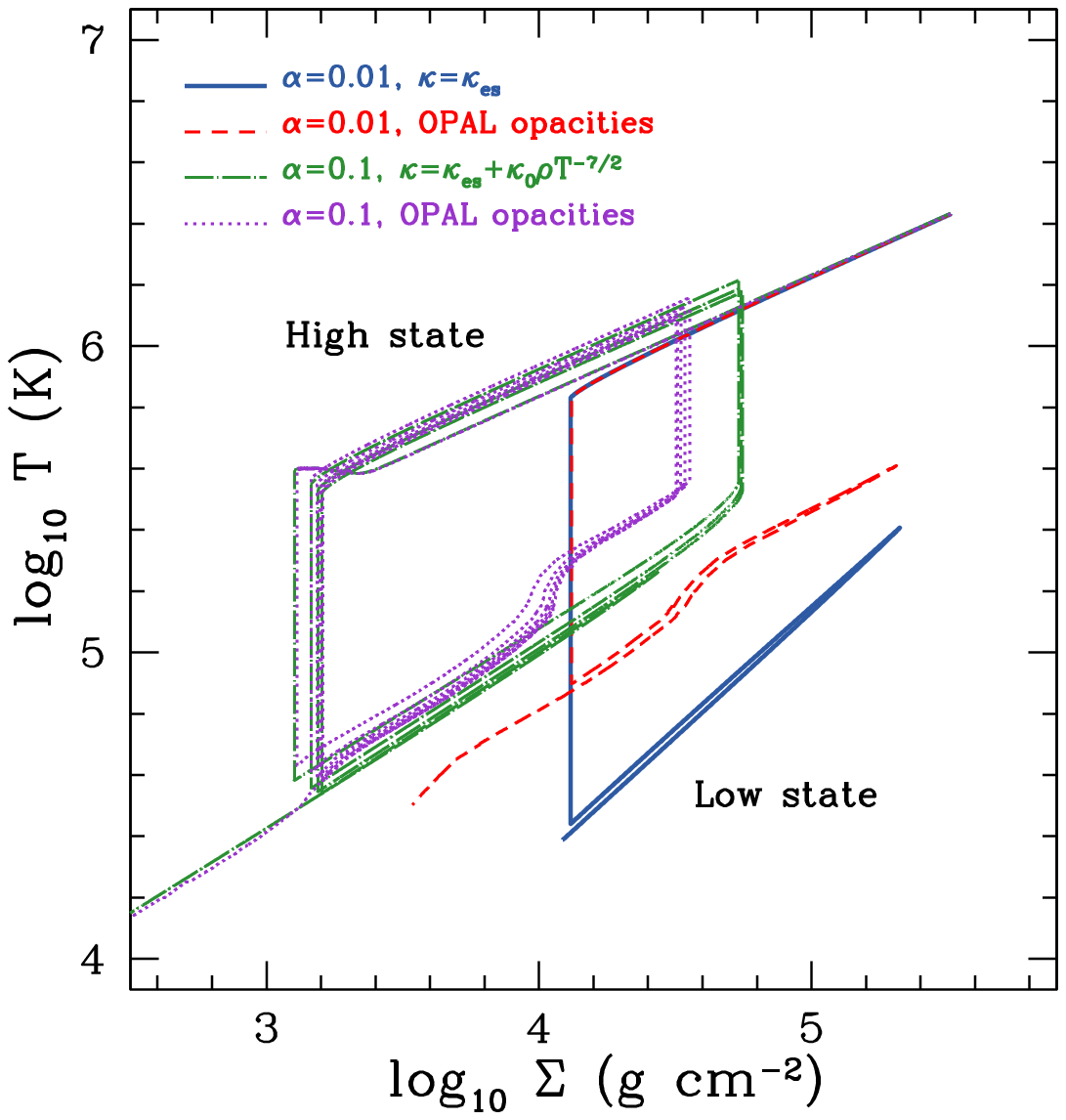}
\caption{The same disk evolution solutions as shown in Figure~\ref{fig:comparison}, but instead plotted as a function of surface density $\Sigma$ and temperature $T$. This highlights how these solutions change between high and low accretion states as they evolve through the thermal instability.}
\label{fig:comparison_sigma_temp}
\end{figure}

In addition, \citet{Shen14} use the analytic fallback rate given in Equation~(\ref{eq:mdotfb}) rather than the numerical fallback rates we show in Figure~\ref{fig:fallback}. Integrating Equation~(\ref{eq:mdotfb}) from $t=t_{\rm fb}$ to $t=\infty$ results in a mass of $3M_*/10$. They assume that an additional mass of $M_*/5$ circularizes during the time $\sim t_{\rm fb}$, so that basically $\sim t_{\rm fb}\times \dot{M}_{\rm fb}(t_{\rm fb})$ sets the initial mass of the disk. This then sums to give a total fallback mass of $M_*/2$ as is well known.

We compare our calculations to the work of \citet{Shen14} in Figure~\ref{fig:comparison}. We scale the disk radius to $R_c$ and the accretion rate to the Eddington rate
\be
    \dot{M}_{\rm Edd}
    &=& L_{\rm Edd}/c^2
    = \frac{4\pi GM_{\rm BH}}{\kappa_{\rm es} c}
    \nonumber
    \\
    &=&2.6\times10^{-2}M_6\,M_\odot\,{\rm yr}^{-1},
    \label{eq:eddington}
\ee
where $L_{\rm Edd}$ is the Eddington luminosity, to mimic their Figure~7 as closely as possible. Their ``fiducial model'' (blue solid line) uses $M_{\rm BH}=10^6\,M_\odot$, $M_*=M_\odot$, $\beta=1$, $\alpha=0.01$, and purely $\kappa=\kappa_{\rm es}$ for the opacity (see the upper left panel in their Figure~7). We get an $R_d\propto t^{2/3}$ at early times like \citet{Shen14}. The accretion rate and radius then drop dramatically at $t\approx10^{-0.6}\,{\rm yrs}$ due to the thermal instability. The accretion rate slowly increases as the disk builds from fallback accretion. At late times once $\dot{M}_{\rm fb} \ll \dot{M}$, {the disk obeys the power-law behavior $\dot{M}\propto t^{-19/16}$ and $R_d\propto t^{3/8}$ as derived by  \citet{Cannizzo1990}.}

\begin{figure*}
\includegraphics[width=0.33\textwidth,trim=0.0cm 0.0cm 0.0cm 0.0cm]{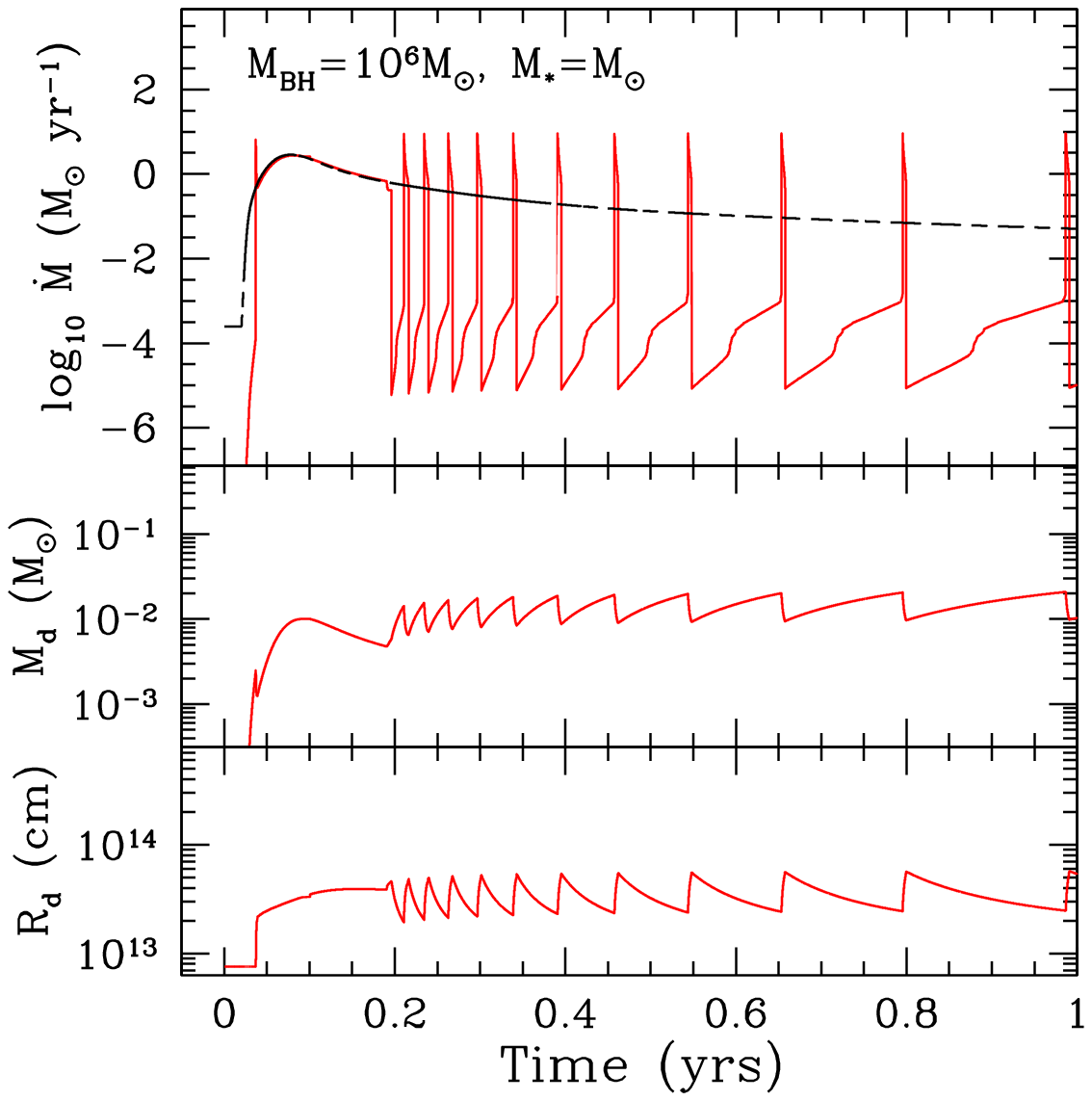}
\includegraphics[width=0.33\textwidth,trim=0.0cm 0.0cm 0.0cm 0.0cm]{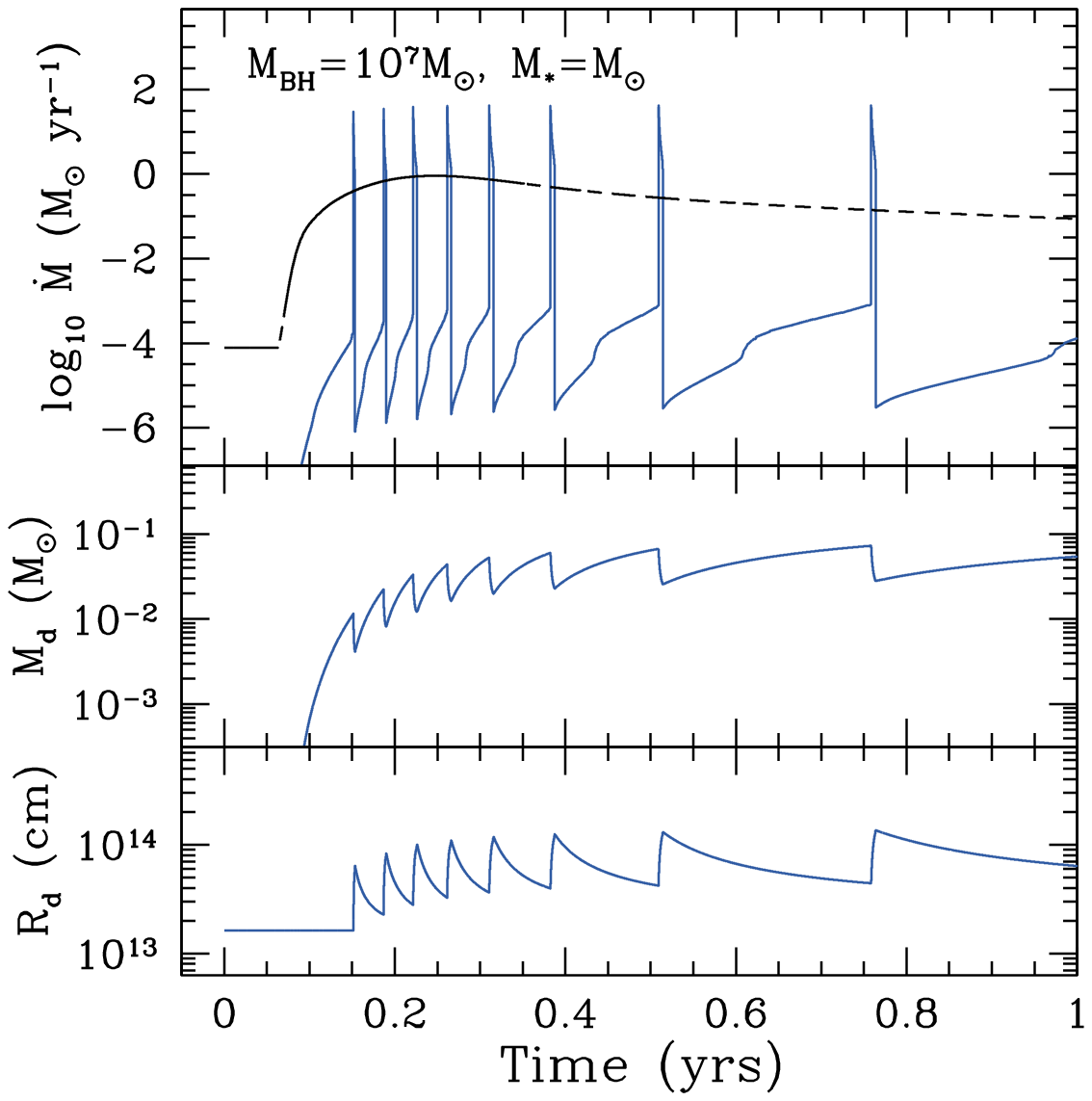}
\includegraphics[width=0.33\textwidth,trim=0.0cm 0.0cm 0.0cm 0.0cm]{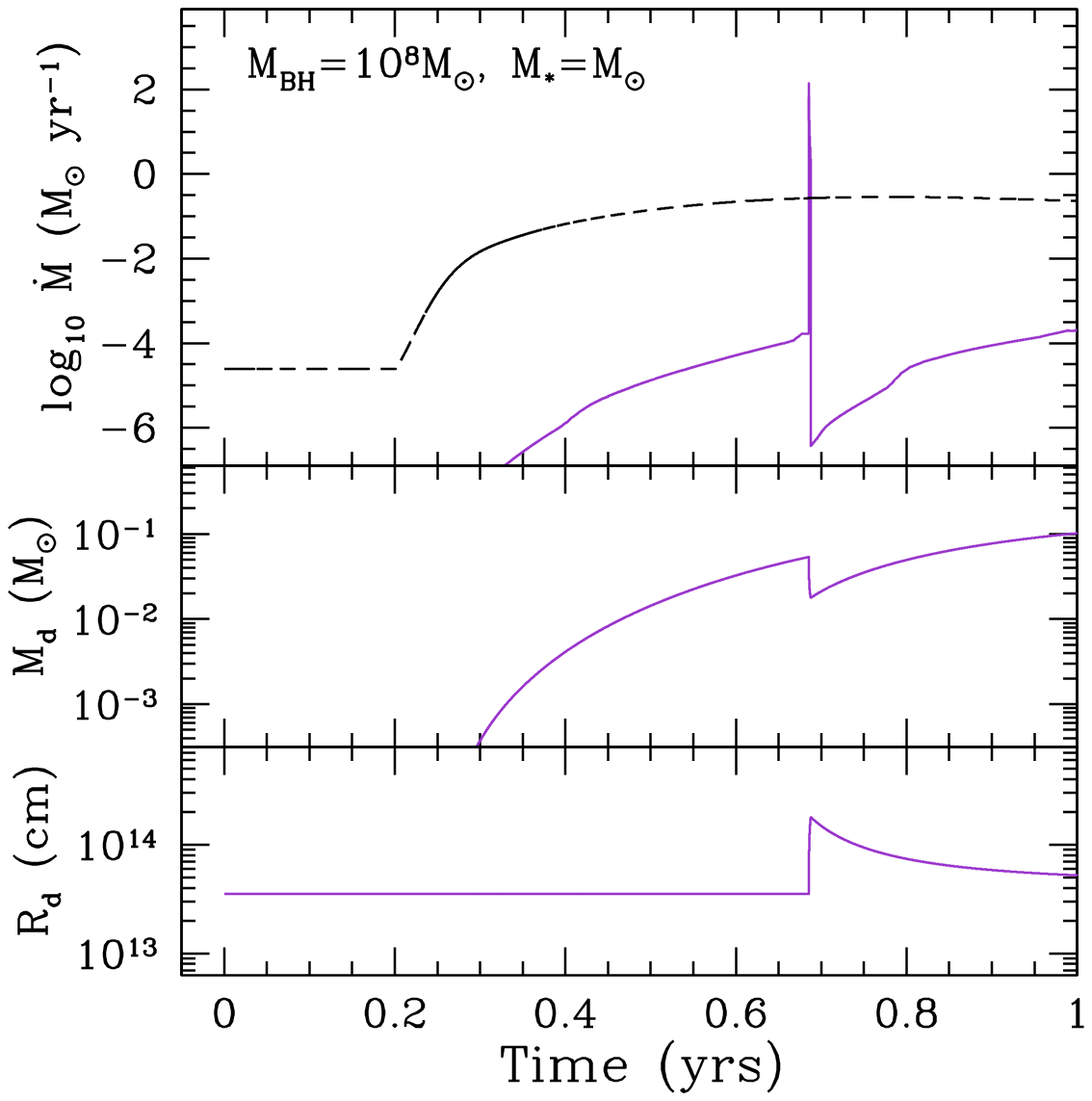}
\caption{Evolution of the accretion rate $\dot{M}$, disk mass $M_d$, and disk radius $R_d$ over the first year using $M_*=M_\odot$, $\beta=1.85$, $\alpha=0.1$, and $p=0.5$ for three different values of $M_{\rm BH}$ as indicated. The black lines delineate the fallback accretion rate $\dot{M}_{\rm fb}$.}
\label{fig:time_evol_short}
\end{figure*}

The other models in Figure~\ref{fig:comparison} explore what happens as we change the model parameters away from the fiducial values used by \citet{Shen14}. First, we use OPAL opacities rather than strictly electron scattering (red dashed lines). The evolution is mostly the same, with the main differences being the late-time radius and an increased $\dot{M}$ during the low state. Next, we increase the viscosity to $\alpha=0.1$ and use the analytic opacity from Equation~(\ref{eq:opacity}) (green dot-dashed lines). The higher viscosity makes the disk evolve more quickly, increasing the disk radius at early times and then causing the disk to go through multiple cycles of low and then high states for a few years until $\dot{M}_{\rm fb} \lesssim \dot{M}$. We note that such cycles were also seen in a subset of models explored by \citet{Shen14}, although this was not the focus on their work. Finally, we use $\alpha=0.1$ with the full OPAL opacities (purple dotted line). Although the high states now appear at different times in comparison to the previous model, the evolution is qualitatively similar. An interesting detail is that with the OPAL opacities $\dot{M}$ is slightly larger as it rises toward a high state.

To further explore the differences between these models, we plot them as a function of $\Sigma$ and $T$ in Figure~\ref{fig:comparison_sigma_temp}. This helps us better focus on how the different models evolve through the thermal instability (as seen from Figure~\ref{fig:sigma_temp_diagram}). Initially, all four models start with a similar $\Sigma$ and $T$ at the top-right corner of the plot, and then evolve toward the bottom left. (Unlike in Figure~\ref{fig:sigma_temp_diagram}, these models start at high accretion rates because of the large initial disk mass that is assumed.) The lower $\alpha$ models (blue solid and red dashed lines) hit the instability at larger $\Sigma$ and $T$ (but not necessarily earlier in time since the viscous time is also controlled by $\alpha$), and quickly drop to the low state. They then climb to the right due to continued feeding from fallback accretion, but they never gain quite enough mass to transition back to the high state and instead eventually trace back toward the left again. One can see that the increased OPAL opacities dramatically raises the position of the low state for these models.

The two high $\alpha$ models (green dot-dashed and purple dotted lines) evolve fairly similarly with the high state extending to somewhat lower temperatures before falling to the low state and now having an even hotter low state due to the higher $\alpha$ value. The important difference here is due to the iron-opacity bump, which increases the temperature and shortens the length of the low state.

\section{Exploring the Disk Solutions}
\label{sec:disksolutions}

\begin{figure*}
\includegraphics[width=0.33\textwidth,trim=0.0cm 0.0cm 0.0cm 0.0cm]{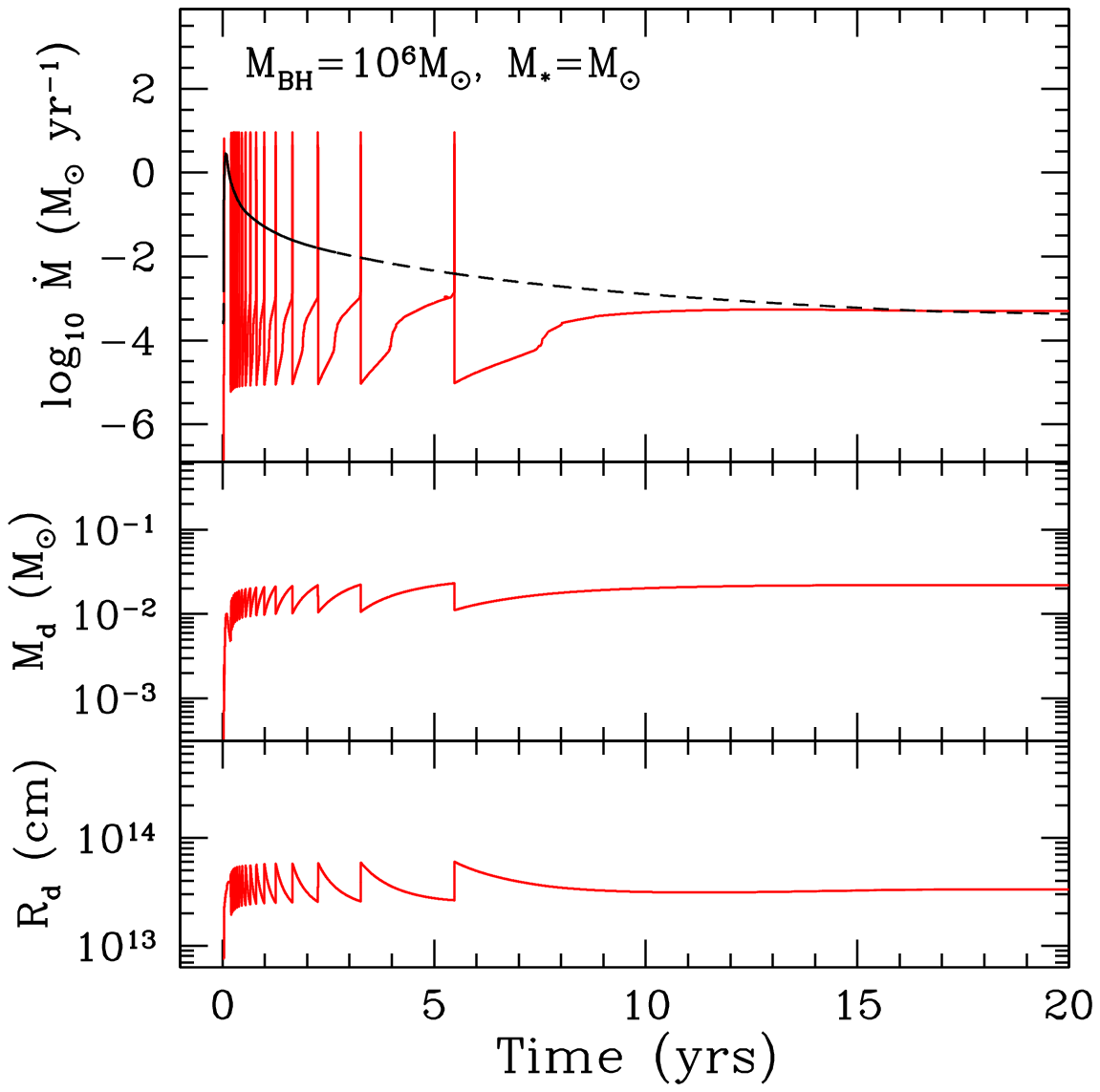}
\includegraphics[width=0.33\textwidth,trim=0.0cm 0.0cm 0.0cm 0.0cm]{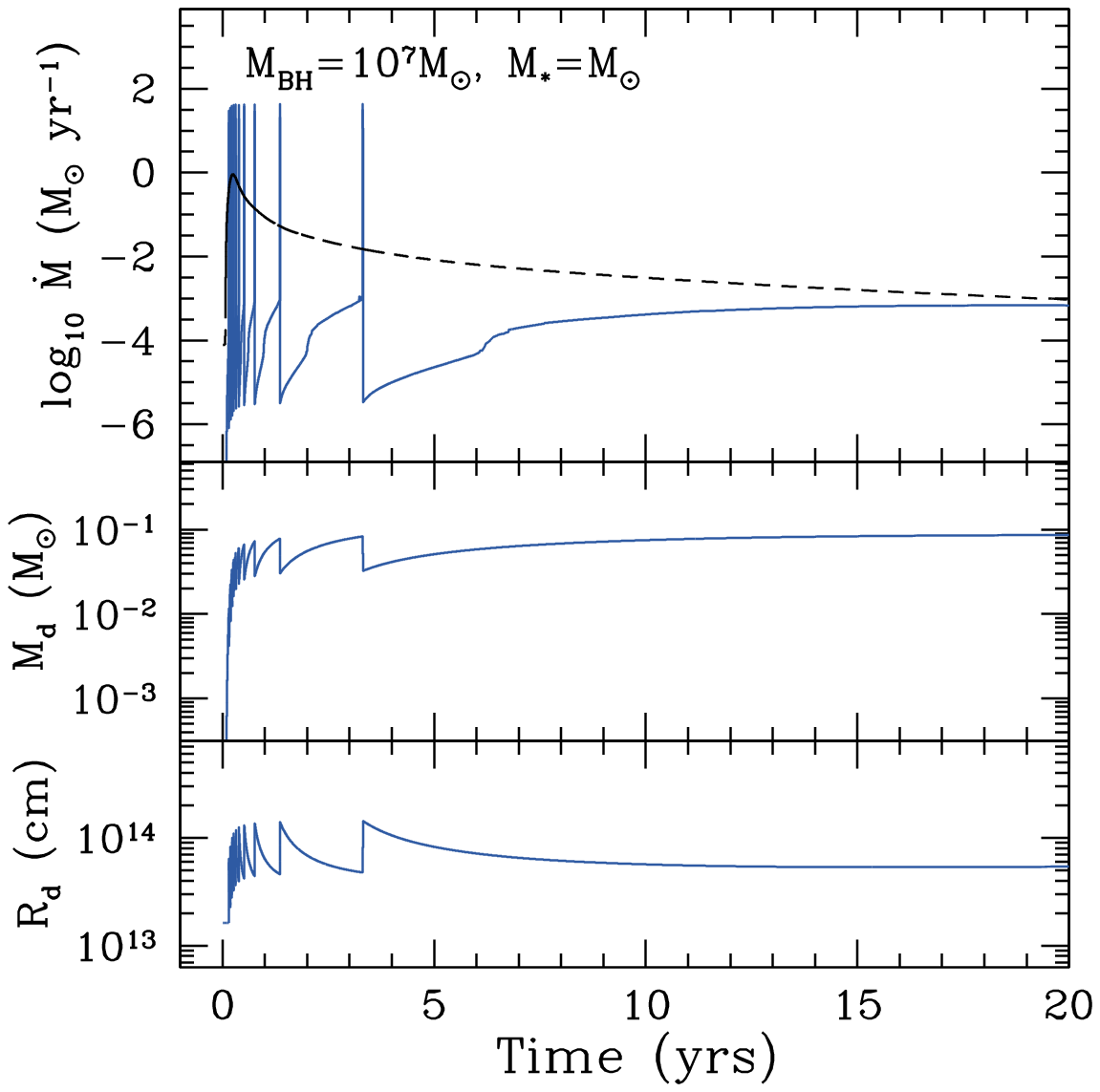}
\includegraphics[width=0.33\textwidth,trim=0.0cm 0.0cm 0.0cm 0.0cm]{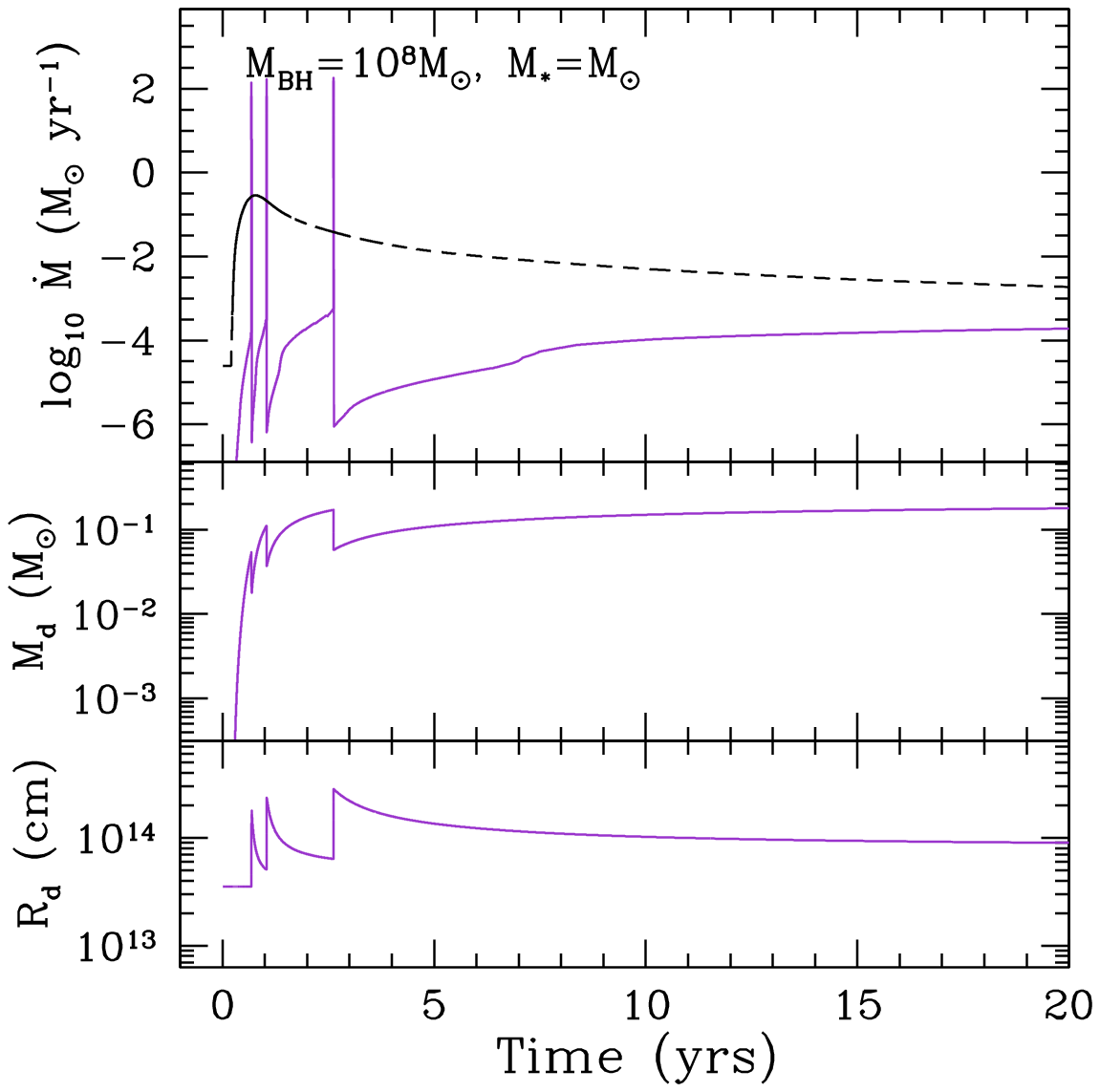}
\caption{The same models as in Figure~\ref{fig:time_evol_short}, but plotted over 20~years instead.}
\label{fig:time_evol_long}
\end{figure*}

Now that we have confirmed our methods and explained the differences between our results and previous work, we return to the prescriptions described in Section~\ref{sec:diskmodel}, along with the numerical fallback rates presented in Figure~\ref{fig:fallback}. We calculate a suite of disk evolutions with time and summarize some of the main results here.

\subsection{Dependence on BH Mass}

In Figure~\ref{fig:time_evol_short}, we consider three BH masses and focus on the first year of the evolution to highlight the features present early on. It can be seen that during each outburst the disk mass decreases dramatically due to the increased $\dot{M}$. The disk radius correspondingly increases to conserve angular momentum. Then, during the low state, the accretion rate grows as the mass of the disk builds from fallback accretion. The radius during the low state shrinks because the fallback material has less specific angular momentum than the disk material. The BH mass plays an important role in setting the viscous time in the disk. For a larger $M_{\rm BH}$, the cycling between high and low states is much slower, the disk mass $M_d$ is generally higher, and the disk radius $R_d$ gets pushed to larger values.

The early evolution also changes for different $M_{\rm BH}$ values. For $M_{\rm BH}=10^6\,M_\odot$, the disk accretion rate closely follows the fallback rate, while for $M_{\rm BH}=10^7\,M_\odot$ or $10^8\,M_\odot$ the accretion rate is well below the fallback rate except during the high states. The reason for this is that the fallback evolution is slow for high mass BHs, so accretion prevents the disk from building at early times. Whether or not this happens for real TDEs likely depends on the details of the circularization process and when a viscous accretion disk is actually established. It may also depend on whether heating from the fallback stream is important (as discussed in Section~\ref{sec:fallbackheating}). For all these reasons, we are hesitant to too strongly interpret this early evolution without using a starting point that is more closely set by detailed simulations. Nevertheless, these early uncertainties do not impact the flaring activity that is exhibited later on, which are insensitive to the initial conditions.

Figure~\ref{fig:time_evol_long} shows the same models as in Figure~\ref{fig:time_evol_short}, but now on a timescale of 20~years to highlight the longer-term evolution. We see that the waiting time between high states increases with time. This is due to the lower fallback accretion rates, which causes the disk to cycle more slowly between low and high states. The fallback rate continues to drop, but is still high enough to power outburst cycles out to almost a decade.

\subsection{Mass Ejections During the High State}

\begin{figure}
\includegraphics[width=0.45\textwidth,trim=0.5cm 0.3cm 1.5cm 0.0cm]{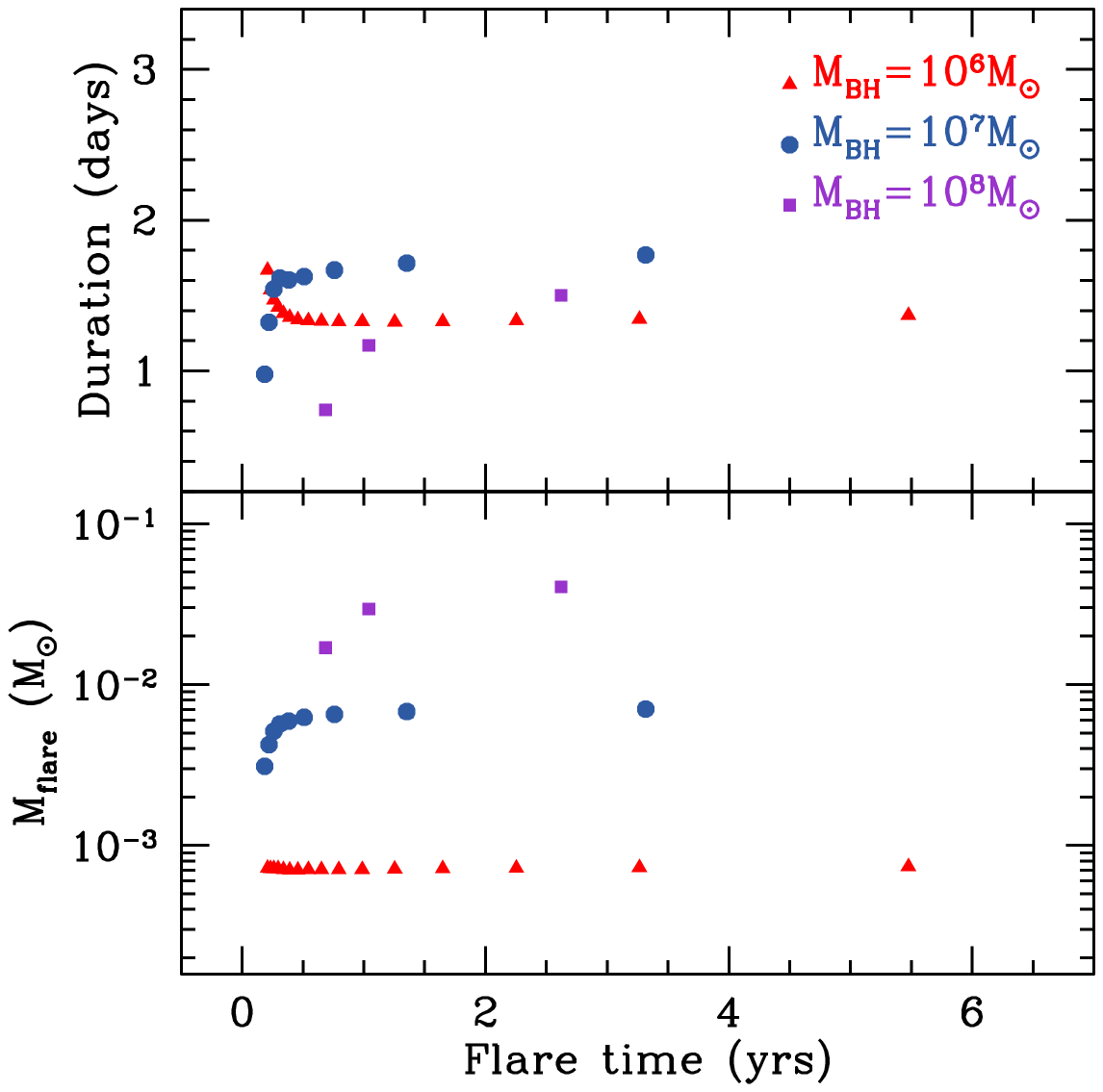}
\caption{The duration of the high state (top panel) and estimate of mass ejected in a flare (bottom panel) using the models from Figures~\ref{fig:time_evol_short} and \ref{fig:time_evol_long}. Different mass BHs are designated different symbols and colors as indicated.}
\label{fig:mej}
\end{figure}

The high states of these models all exceed the Eddington accretion rate, Equation (\ref{eq:eddington}), thus there is probably heavy mass loss during these phases \citep[e.g.,][]{Blandford1999,Dai2018,Thomsen2022}. In Figure~\ref{fig:mej}, we plot the duration of the high states in the upper panel and an estimate of the mass lost, which is the integral
\be
    M_{\rm flare}
    =\int \dot{M}_{\rm out}(t)dt,
\ee
using Equation~(\ref{eq:mdotout}). This shows that in principle the high state flares can occur out to $\lesssim6\,{\rm yrs}$ after the TDE. The typical duration is $\sim1-2\,{\rm days}$, which is fairly insensitive to the BH mass. The mass ejected during this phase $M_{\rm flare}\sim10^{-3}-10^{-1}\,M_\odot$, with larger BHs ejecting more mass per flare.

\subsection{Changes with $M_*$, $\alpha$, and $p$}

\begin{figure*}
\includegraphics[width=0.33\textwidth,trim=0.0cm 0.0cm 0.0cm 0.0cm]{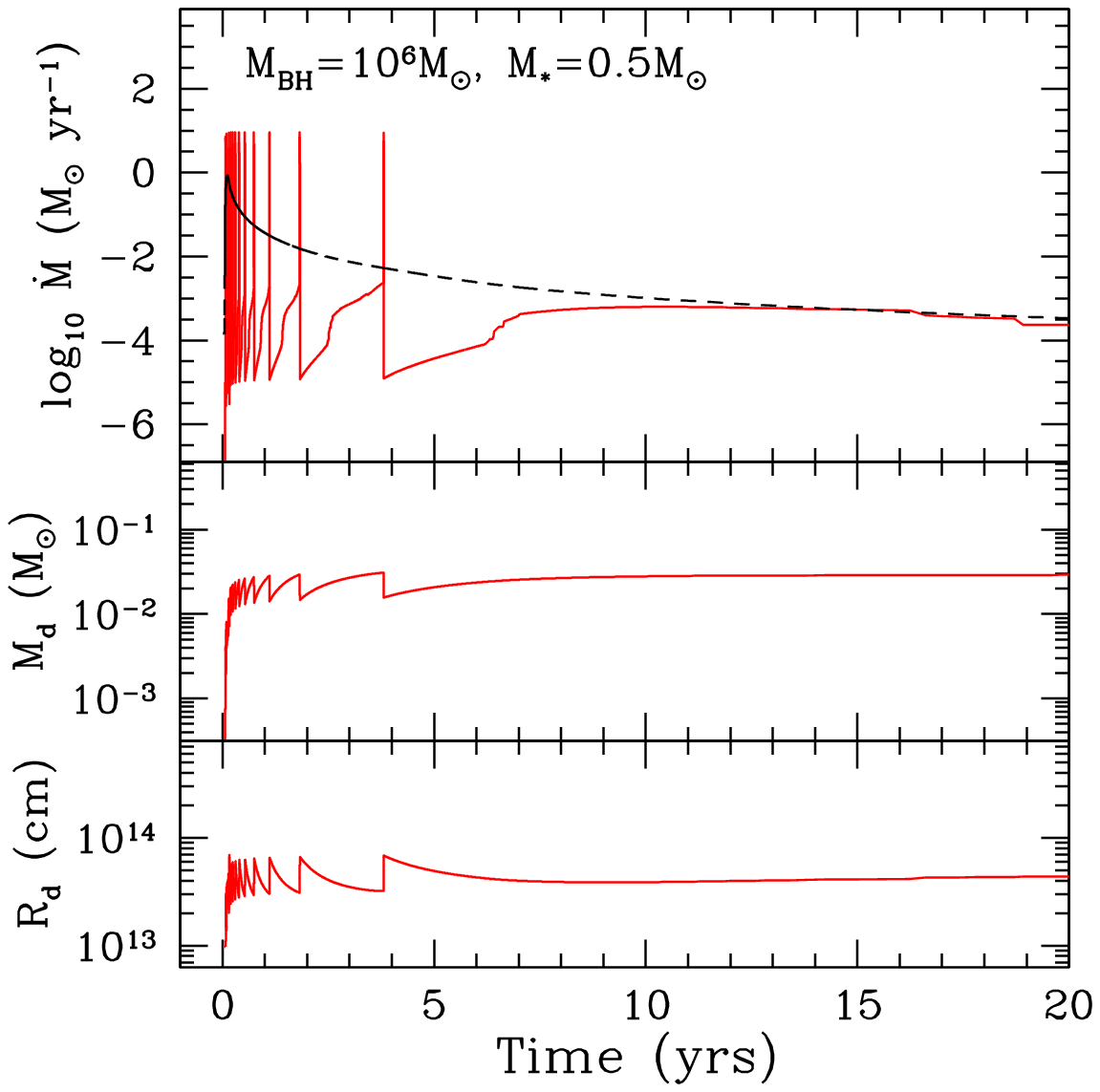}
\includegraphics[width=0.33\textwidth,trim=0.0cm 0.0cm 0.0cm 0.0cm]{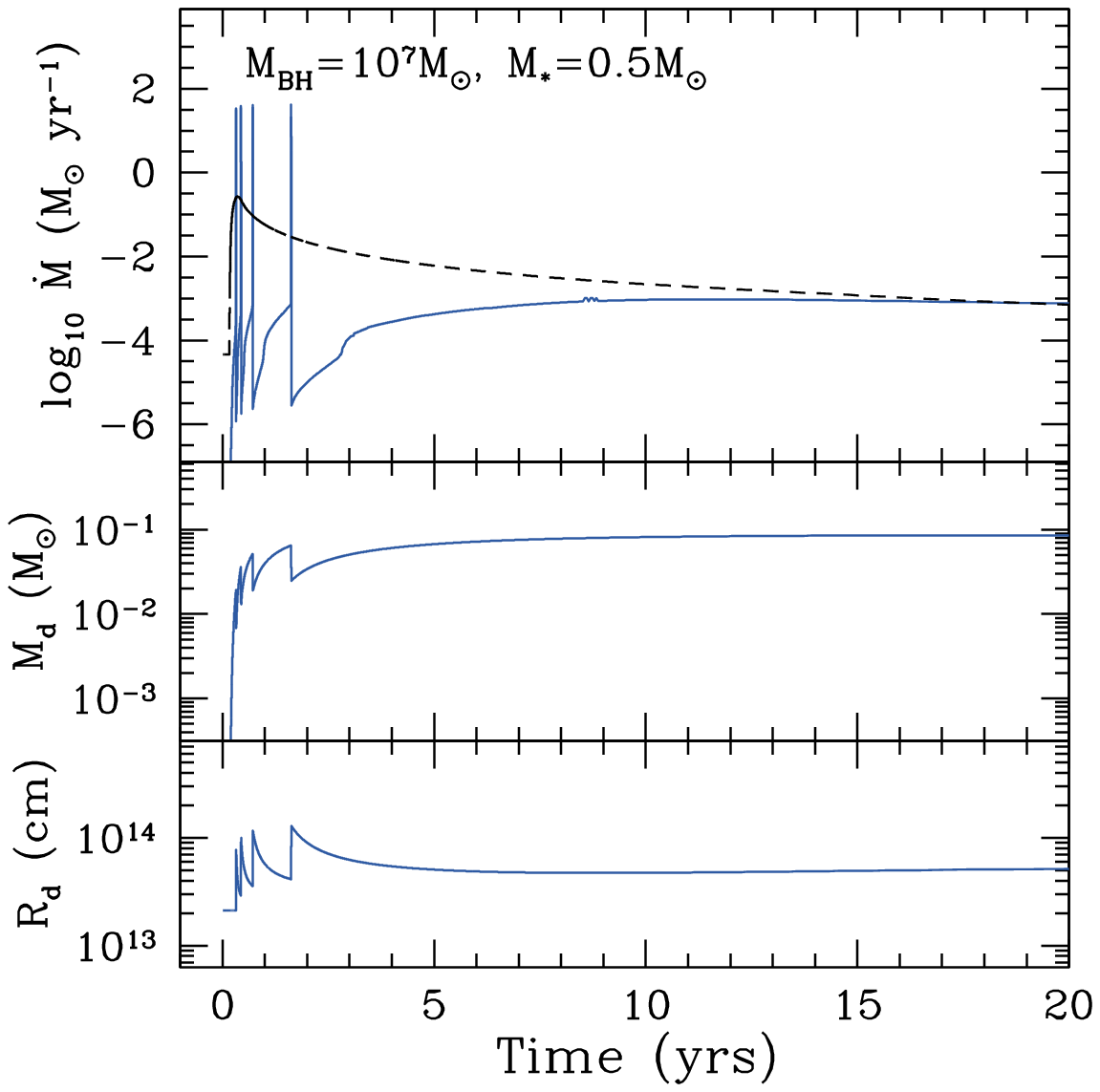}
\includegraphics[width=0.33\textwidth,trim=0.0cm 0.0cm 0.0cm 0.0cm]{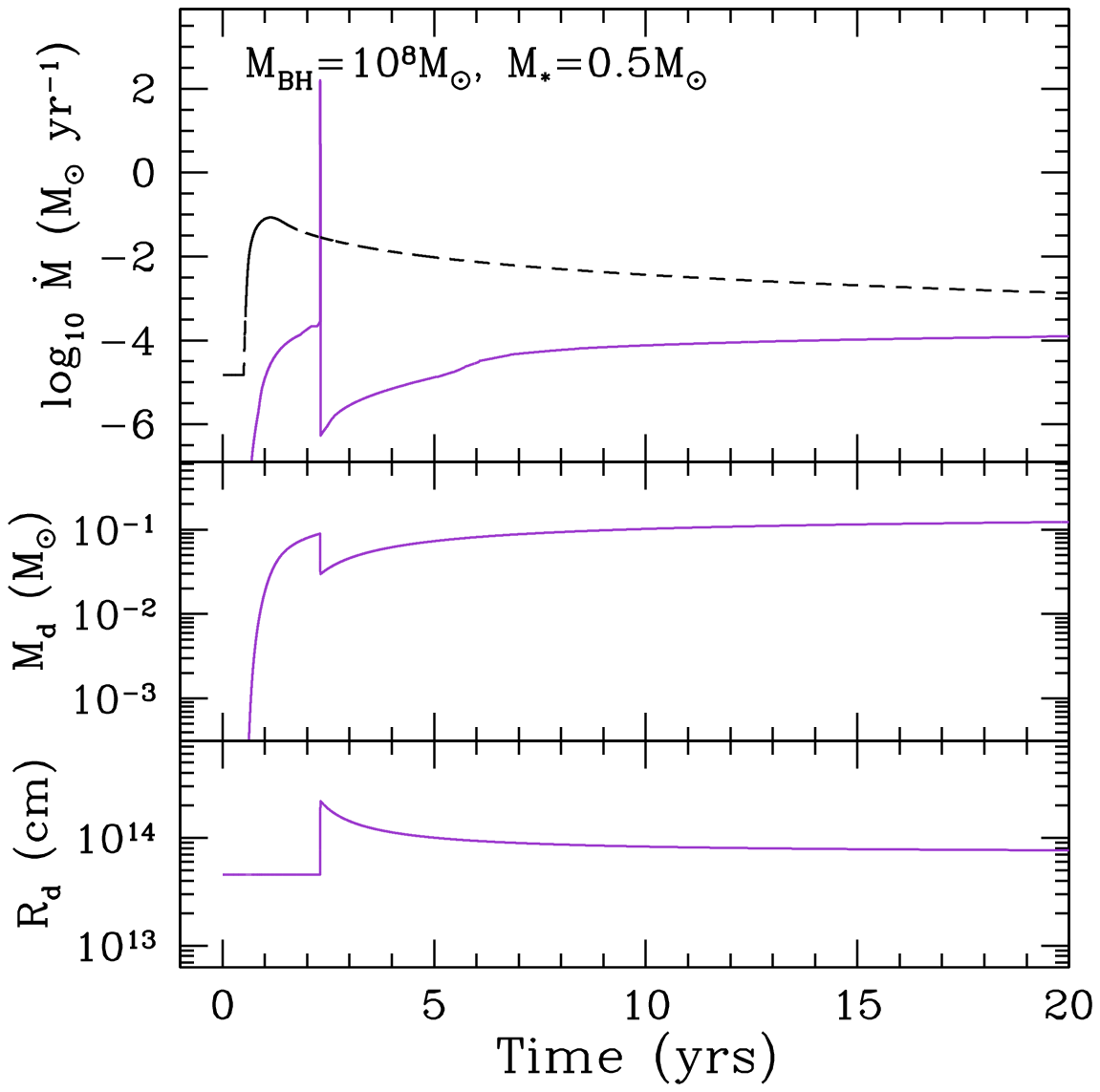}
\caption{The same as Figure~\ref{fig:time_evol_long}, but using $M_*=0.5\,M_\odot$ and $\beta=0.9$ instead.}
\label{fig:time_evol_lowmass}
\end{figure*}

In Figure~\ref{fig:time_evol_lowmass}, we explore how the time-dependent evolution changes for a lower mass star in the TDE. We set $M_*=0.5\,M_\odot$ and use the fallback rate for an $n=5/3$ polytrope since it is more appropriate for a main-sequence star of this mass. We also set $\beta=0.9$ since this is $\beta_c$ for a less centrally concentrated star. Overall, the evolution between high and low states is slower with less high states. The general $\dot{M}$, $M_d$ and $R_d$ values are actually not that different between $M_*=M_\odot$ and $M_*=0.5\,M_\odot$ because the fallback rates are not really that different after peak (as highlighted by Figure~\ref{fig:fallback}).

\begin{figure*}
\includegraphics[width=0.33\textwidth,trim=0.0cm 0.0cm 0.0cm 0.0cm]{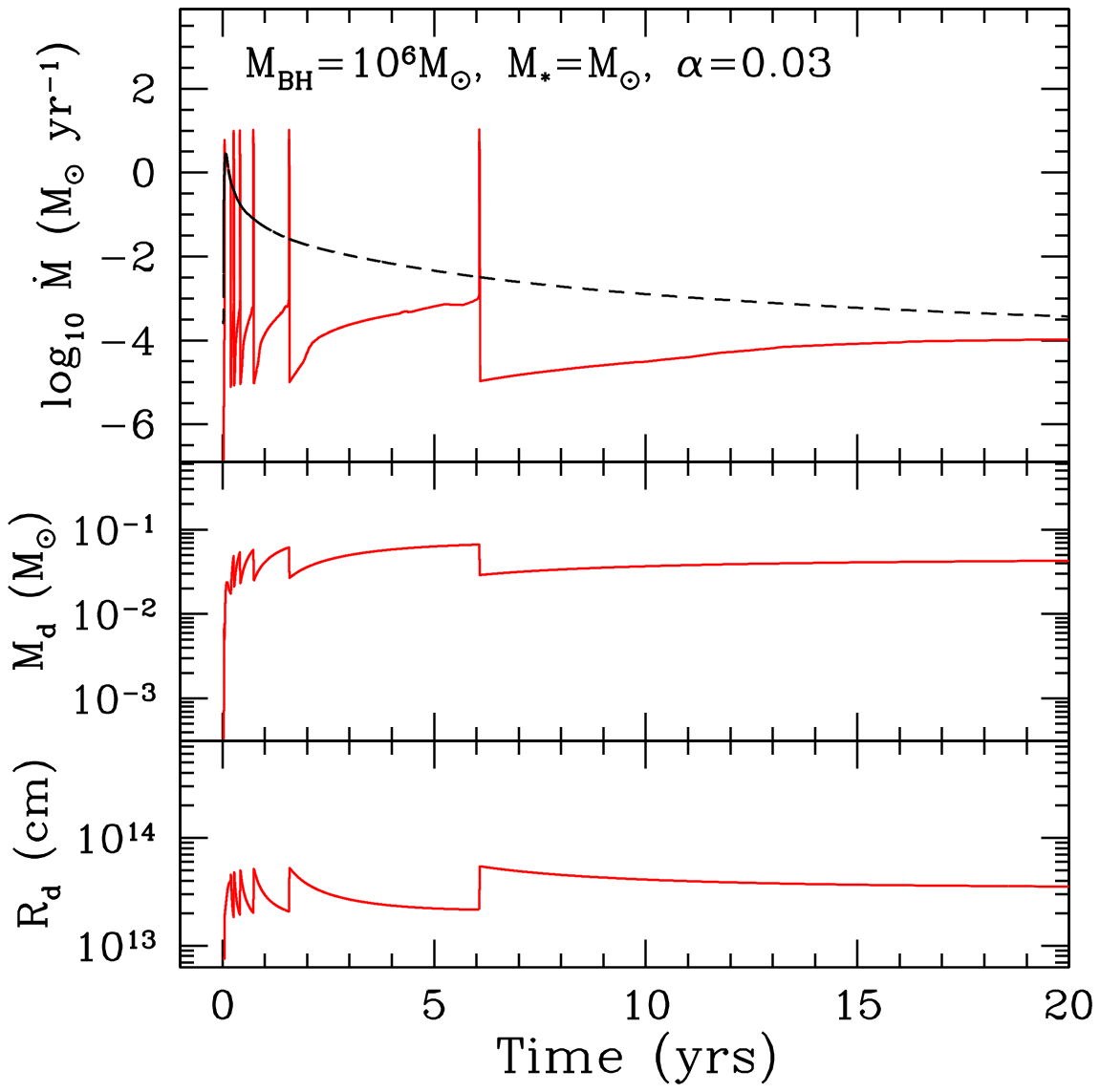}
\includegraphics[width=0.33\textwidth,trim=0.0cm 0.0cm 0.0cm 0.0cm]{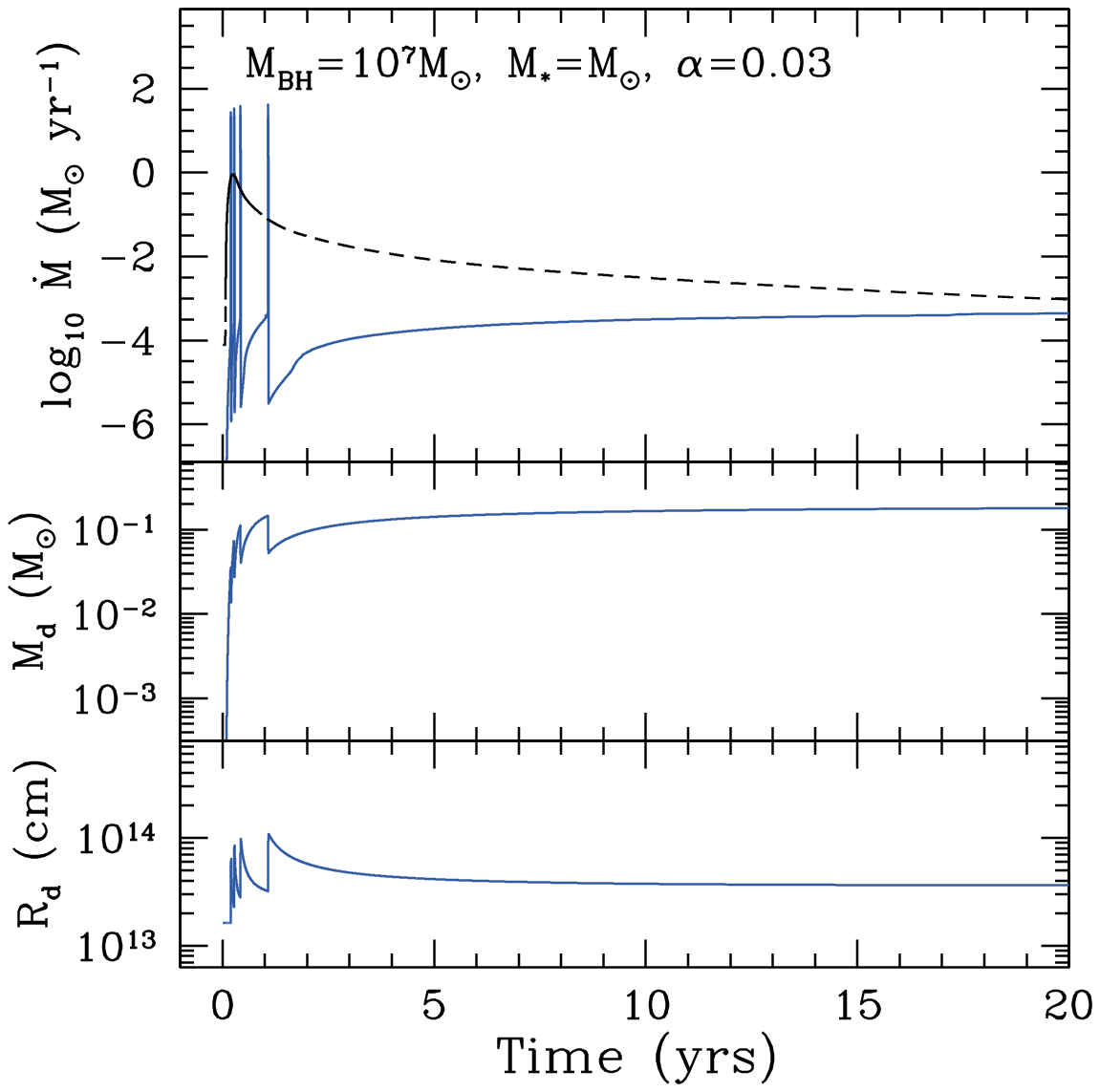}
\includegraphics[width=0.33\textwidth,trim=0.0cm 0.0cm 0.0cm 0.0cm]{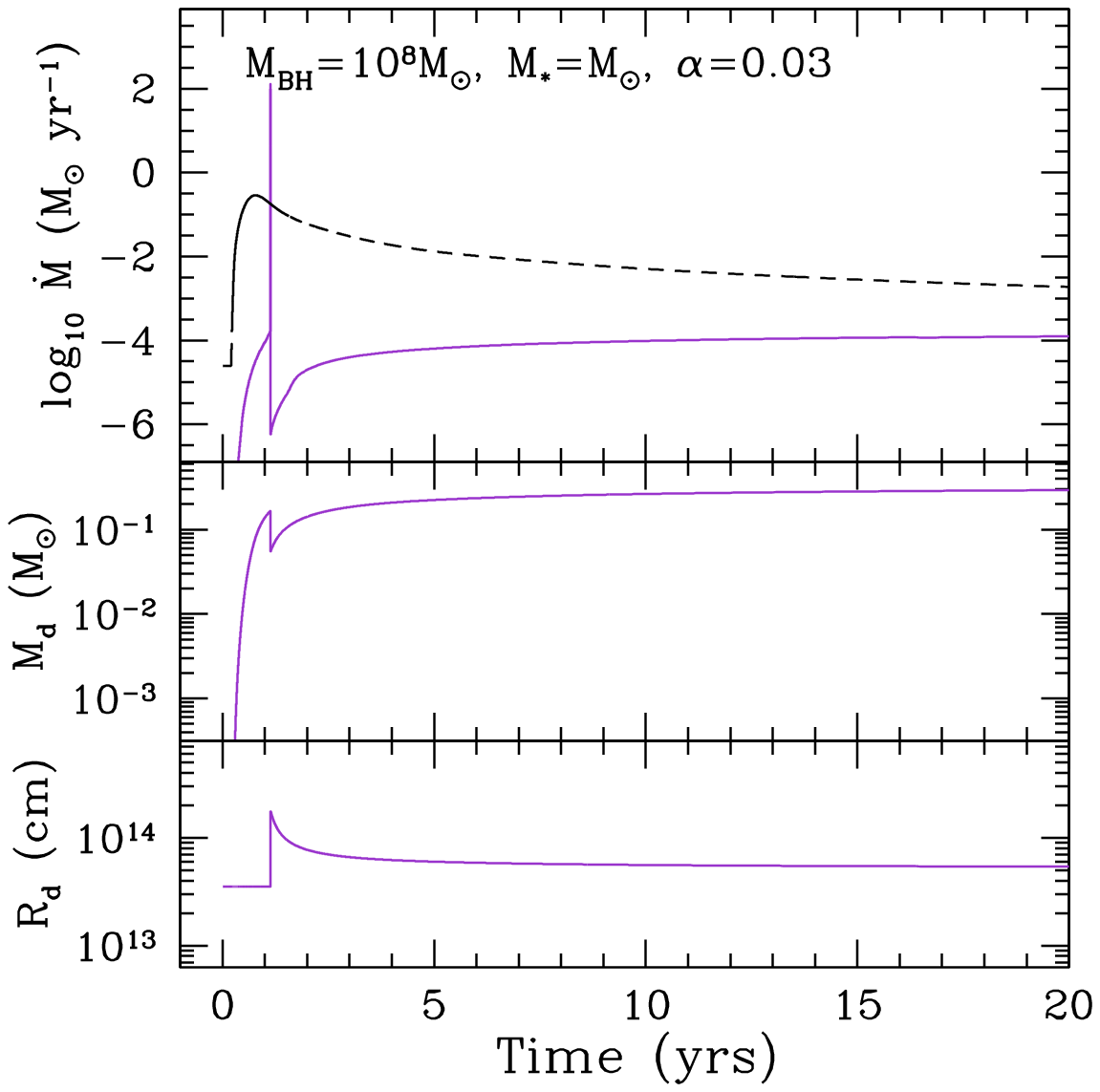}
\caption{The same as Figure~\ref{fig:time_evol_long}, but using $\alpha=0.03$ instead.}
\label{fig:time_evol_lowalpha}
\end{figure*}

In Figure~\ref{fig:time_evol_lowalpha}, we rerun the simulations with $\alpha=0.03$, corresponding to a lower disk viscosity. Again, this causes the cycling between high and low states to be slower with many fewer high states. The actual values of $\dot{M}$, $M_d$ and $R_d$ do not change that dramatically in the low state between $\alpha=0.1$ and $\alpha=0.03$ because these are mostly set by the physics at the transitions between states. This provides some robustness to these models in the sense that these outbursts should happen even if we are not able to predict the exact times when the outbursts will occur.

\begin{figure*}
\includegraphics[width=0.33\textwidth,trim=0.0cm 0.0cm 0.0cm 0.0cm]{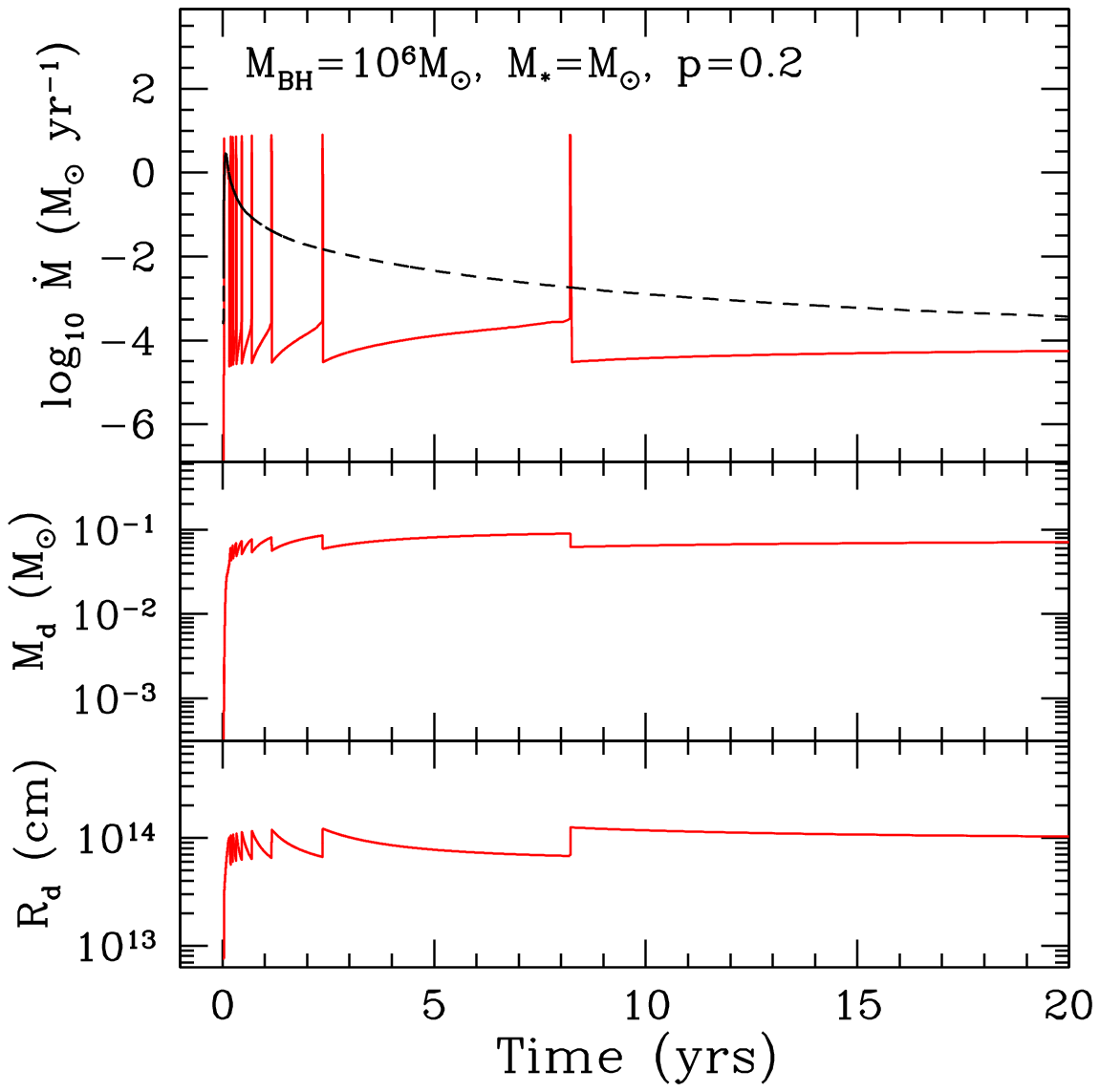}
\includegraphics[width=0.33\textwidth,trim=0.0cm 0.0cm 0.0cm 0.0cm]{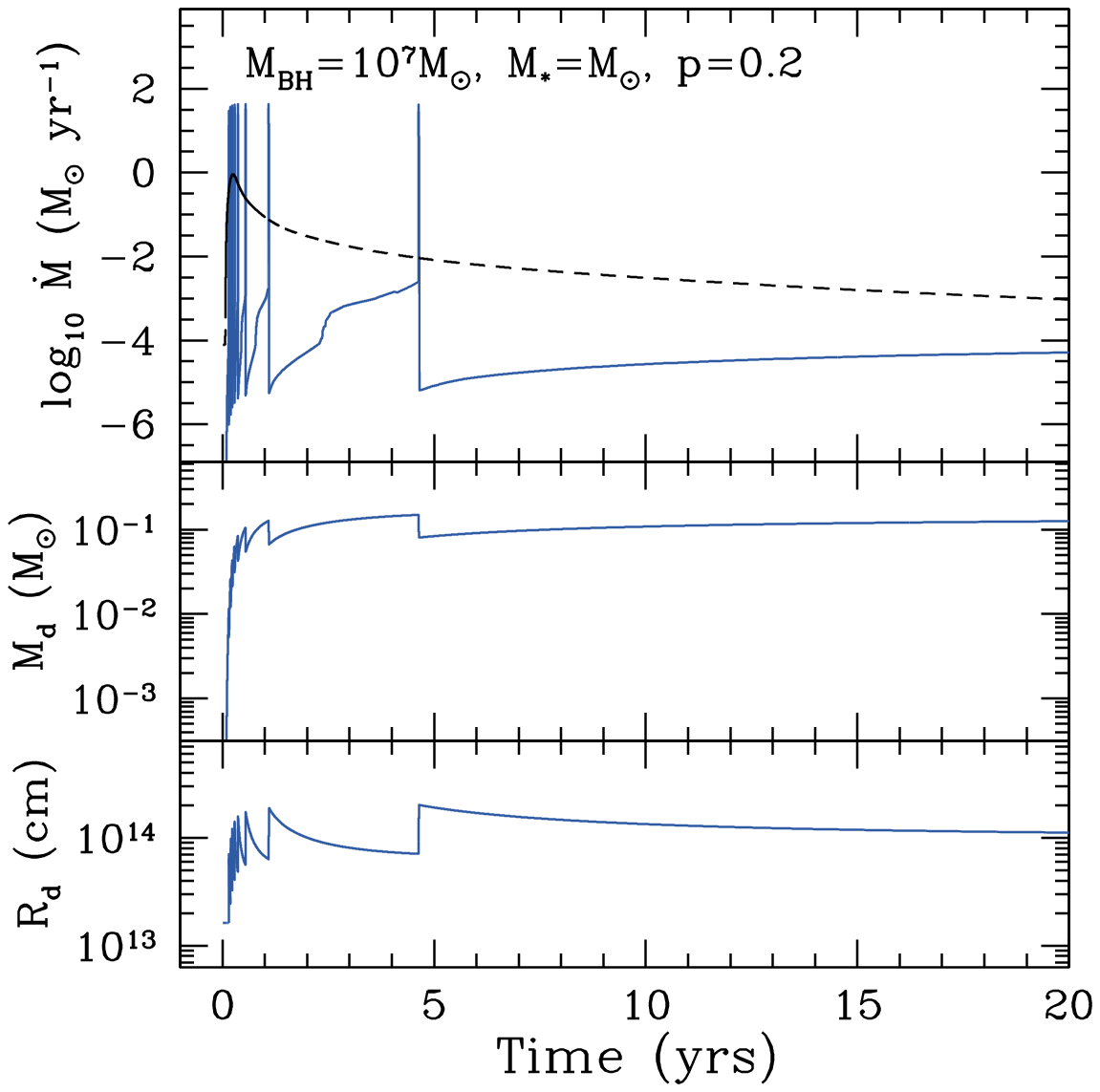}
\includegraphics[width=0.33\textwidth,trim=0.0cm 0.0cm 0.0cm 0.0cm]{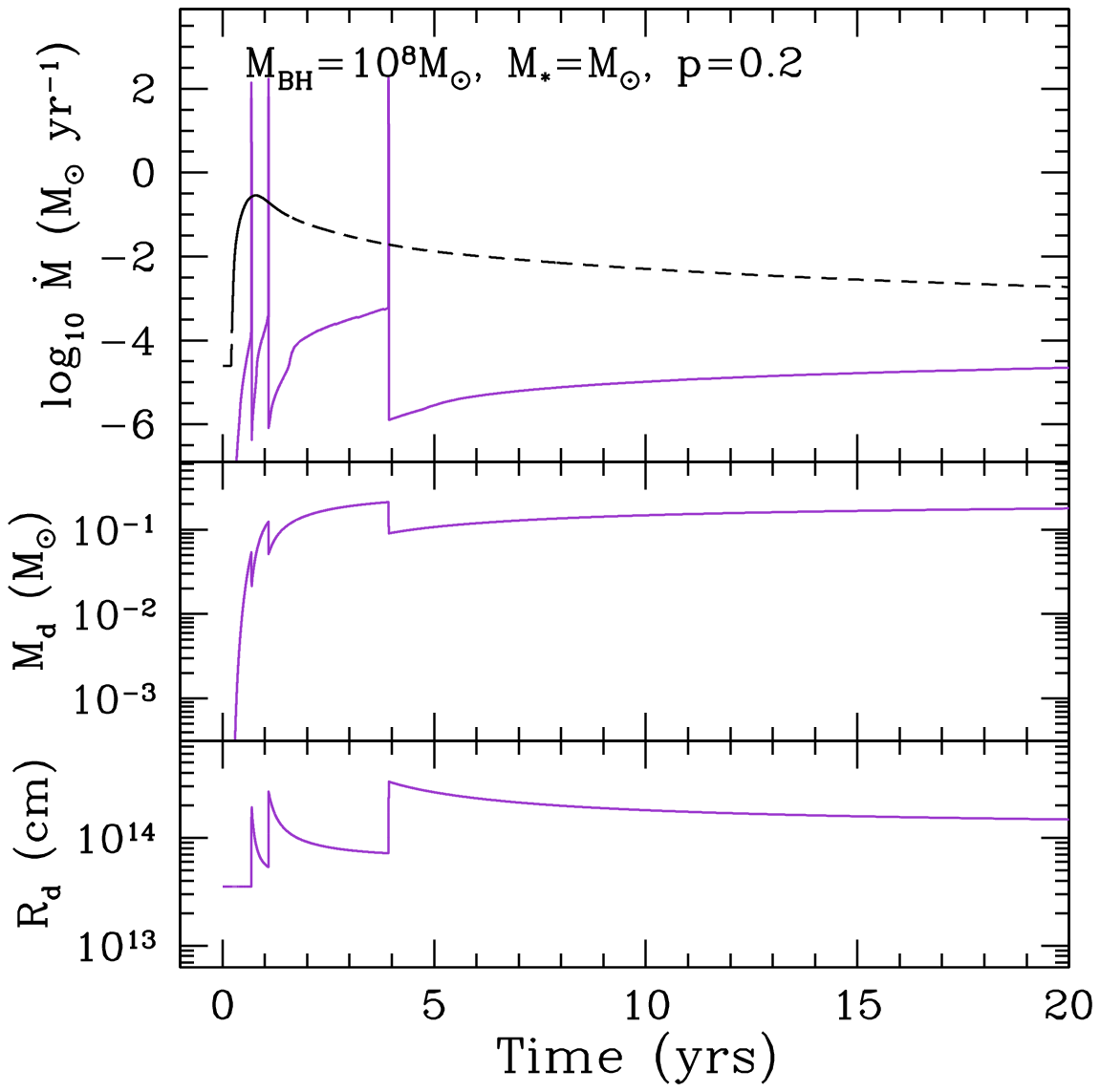}
\caption{The same as Figure~\ref{fig:time_evol_long}, but using $p=0.2$ instead.}
\label{fig:time_evol_lowp}
\end{figure*}

Finally, in Figure~\ref{fig:time_evol_lowp}, we rerun the simulations with $p=0.2$, corresponding to less mass loss in the high state. The main impact is that the disk grows to larger sizes due to the decreased loss of angular momentum. This in turn increases the viscous time so that the disk evolve thought less outburst cycles.

\section{Comparisons to Observations}
\label{sec:observations}

As discussed in Section~\ref{sec:intro}, some of the evidence for long-lasting accretion comes in the form of late-time optical/UV emission and delayed radio flares. We thus compare our disk models with these observations here.

\subsection{Late-time UV Emission}

Assuming the disk radiates as a series of black bodies at each annulus, the effective temperature at a radius $r$ is
\be
    T_{\rm eff} (r)
    = \left\{
        \frac{3GM_{\rm BH}\dot{M}}{8\pi\sigma_{\rm SB}r^{3}}
        \lb 1-\lp \frac{R_i}{r}\rp\rb^{1/2}
    \right\}^{1/4},
\ee
where $\sigma_{\rm SB}$ is the Stefan-Boltzmann constant. {Radiative transfer effects can complicate the relationship between the observed temperature and $T_{\rm eff}$ \citep[e.g.][]{Done2012}, but since we are only doing simply comparisons at this point we save these details for future work.} Since we are considering BH masses up $10^8\,M_\odot$, the BH spin must be fairly high to allow the TDE to occur. To be consistent, we assume $R_i=R_s/2$ (a maximally spinning BH) for our spectral models. Thus there could be strong differences between what we predict and reality at the shortest wavelengths depending on  the BH spin.

Integrating a Planck function over the entire disk gives the observed flux at a given frequency $\nu$ at a distance $D$ \citep{Frank02},
\be
    F_\nu =\frac{4\pi h \nu^3\cos \theta }{c^2D^2}
    \int_{R_s}^{R_d}\frac{rdr}{\exp[h\nu/k_{\rm B}T_{\rm eff}(r)]-1},
    \label{eq:fnu}
\ee
where $\theta$ is the inclination ($\cos \theta = 1$ for a face-on disk). The total luminosity from a disk at a given frequency is calculated by integrating the disk emission over a sphere at radius $D$,
\be
    L_\nu =2\int_0^{2\pi}\int_0^{\pi/2}D^2 F_\nu\sin\theta d\theta d\phi,
    \label{eq:lnu}
\ee
where the factor of $2$ is for the two sides of the disk. Comparing Equation~(\ref{eq:fnu}) with the result from integrating Equation~(\ref{eq:lnu}), we conclude that the conversion between an observed flux and the total luminosity of a disk is
\be
    L_\nu =\frac{2\pi D^2}{\cos \theta}F_\nu.
\ee
In contrast, an observer calculating the luminosity from an observed flux would typically not take into account the disk's emission pattern and instead simply calculate an isotropic equivalent
\be
    L_\nu^{\rm iso} = 4\pi D^2 F_\nu.
\ee
So for comparisons with observations, we precede in the same way.

\begin{figure}
\includegraphics[width=0.45\textwidth,trim=0.5cm 0.3cm 1.5cm 0.0cm]{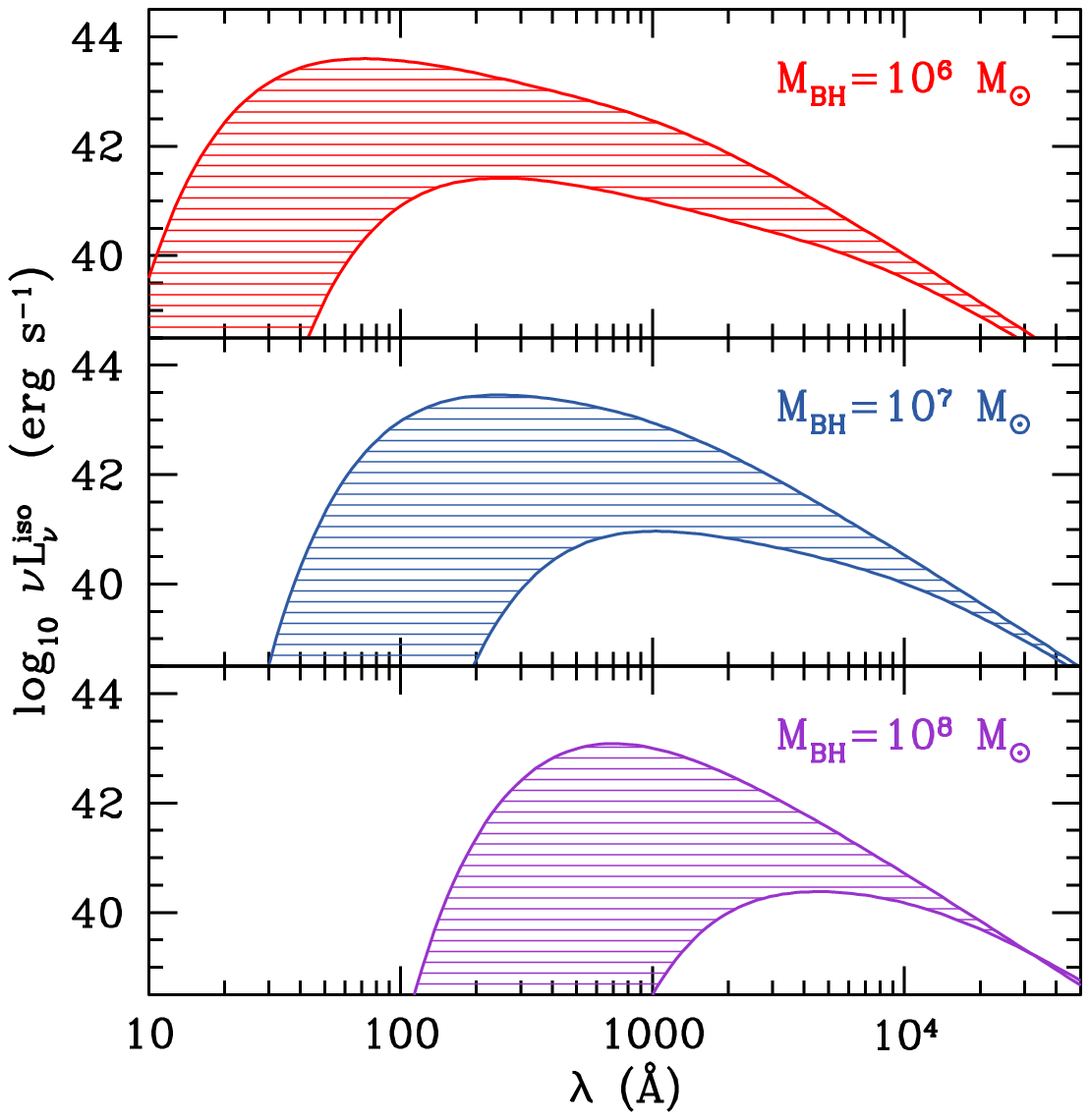}
\caption{Characteristic spectral energy distributions for three different BH masses. In each case, the shaded region represents the range of $\dot{M}$ values possible during the low state (see text for the specific values considered). All models use $M_*=M_\odot$, $\beta=1.85$, $\alpha=0.1$, $p=0.5$, and $\cos \theta =1$.}
\label{fig:sed_multi}
\end{figure}

In Figure~\ref{fig:sed_multi}, we consider the range of spectral energy distributions expected for three different BH masses. In each case we use the parameters from the calculations in Figures~\ref{fig:time_evol_short} and \ref{fig:time_evol_long} ($M_*=M_\odot$, $\beta=1.85$, and $\alpha=0.1$). We assume the viewing angle is face on, so  $\cos \theta =1$. The shaded regions represent the range of accretion rates each BH exhibits while in the low state. For $M_{\rm BH}=10^6\,M_\odot$, this is $9.1\times10^{-6}-1.5\times10^{-3}\,M_\odot\,{\rm yr}^{-1}$, for $M_{\rm BH}=10^7\,M_\odot$, this is $3.4\times10^{-6}-1.0\times10^{-3}\,M_\odot\,{\rm yr}^{-1}$, and for $M_{\rm BH}=10^8\,M_\odot$, this is $8.8\times10^{-7}-5.8\times10^{-4}\,M_\odot\,{\rm yr}^{-1}$. The radius varies as well for each accretion rate, but no more than about a factor of $\sim3$ for a given BH mass. This shows how the overall disk emission can vary dramatically throughout the low state and how a larger BH mass shifts the distribution to longer wavelengths.

\begin{figure}
\includegraphics[width=0.45\textwidth,trim=0.5cm 0.3cm 1.5cm 0.0cm]{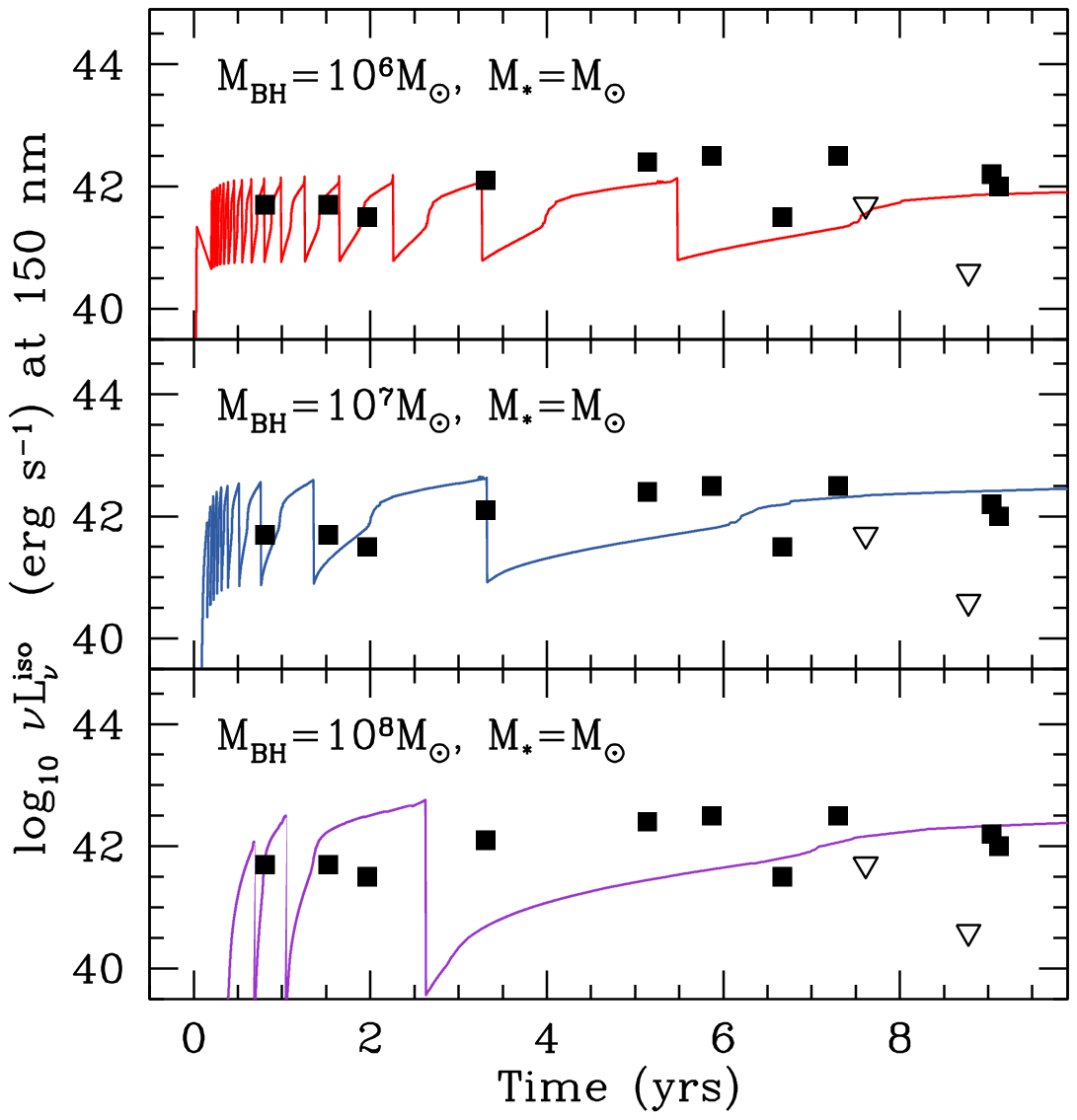}
\caption{Comparison of the isotropic equivalent luminosity at $150\,{\rm nm}$ for 3 different BH masses with the observations summarized in \citet{vanVelzen19}. Detections are shown as filled squares while upper limits are open triangles. Note that this is a combination of observations from 12 different events, and should not be viewed as a light curve (this is further discussed in the text). All models use $M_*=M_\odot$, $\beta=1.85$, $\alpha=0.1$, $p=0.5$, and $\cos \theta =1$.}
\label{fig:uv_comparison}
\end{figure}

In Figure~\ref{fig:uv_comparison}, we compare the time evolution of models with 3 different BH masses to the UV observations summarized in \citet{vanVelzen19}. The models again all use $M_*=M_\odot$, $\beta=1.85$,  $\alpha=0.1$, and a viewing angle of $\cos \theta =1$. The filled squares are UV detections of 10 different TDEs at $295-3332\,{\rm days}$ after the maximum early emission, although we caution that for the first couple of points (at $295$ and $557\,{\rm days}$) there may be significant contribution from a non-disk source such as a reprocessing region in the TDE. We also plot 2 upper limits as open triangles. Since this is combining data from different events, these should not be viewed as a light curve but rather just provides a range of possible luminosities and timescales. We omit plotting the luminosity when the disk is in the high state because the super-Eddington rates may obscure the disk or at least change the disk spectral energy distribution, and anyways, these last for a very short time ($\sim2,{\rm days}$) compared to the overall evolution.

This comparison shows that these models can in principle produce sufficient UV luminosity to explain the observed UV emission. One can distinctly see two components during the low state evolution, which is due to the increased opacity from the iron bump. The two upper limits are also consistent with this picture that the late-time UV for any given system may be dimmer by about an order of magnitude as they evolve through the low state. In the future it will be critical to follow the evolution of TDEs in the UV at higher cadence to better understand the duty cycle of this emission.

\begin{figure}
\includegraphics[width=0.45\textwidth,trim=0.5cm 0.3cm 1.5cm 0.0cm]{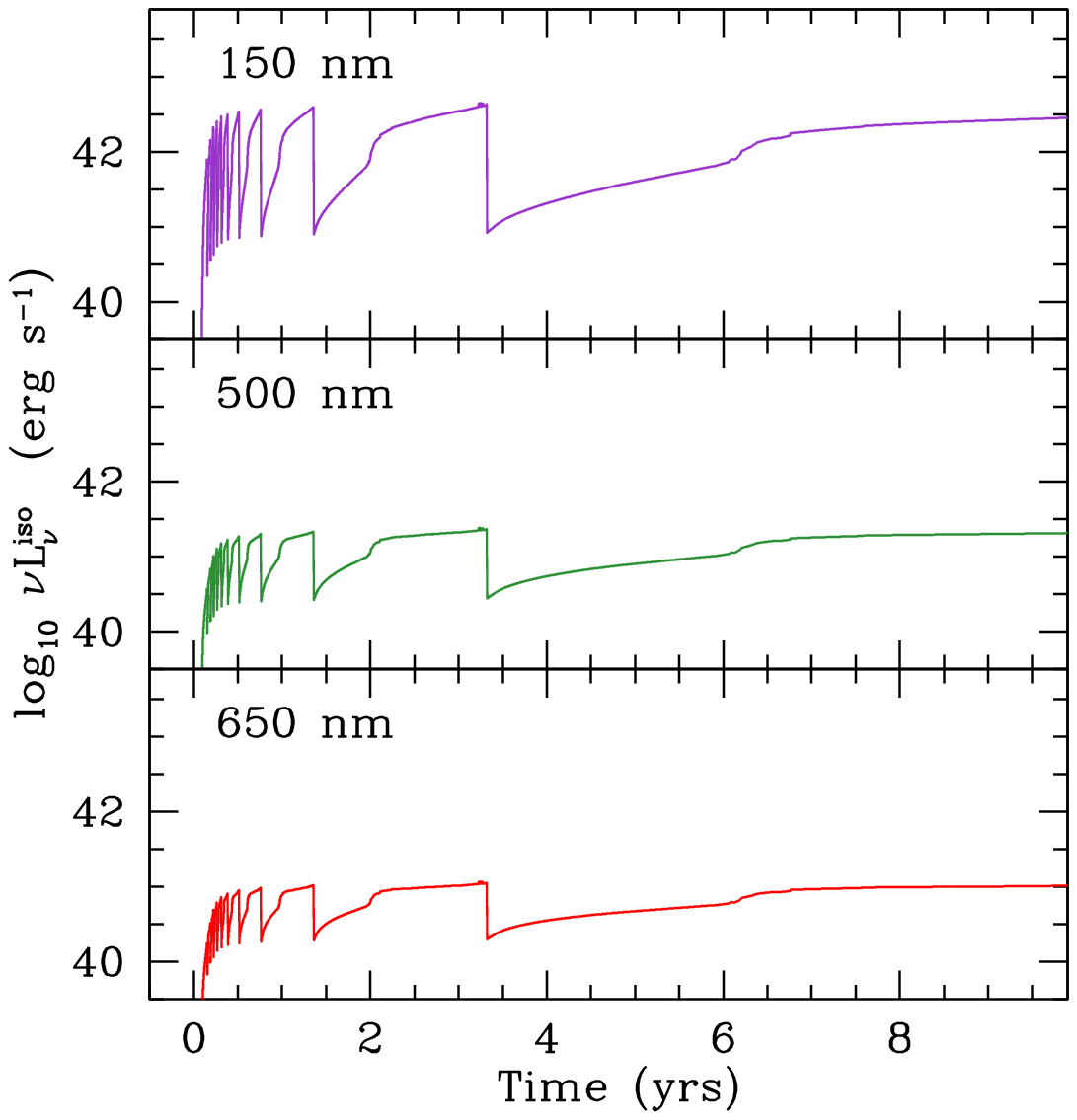}
\caption{Comparison of how a disk light curve may appear in different wavebands using our $M_{\rm BH}=10^7\,M_\odot$ model (middle panel from Figure~\ref{fig:uv_comparison}). The wavelength of emission is noted in each panel. This demonstrates that the flux variations decrease with longer wavelengths.}
\label{fig:optical_comparison}
\end{figure}

From Figure~\ref{fig:sed_multi}, it is apparent that the amount of variation in luminosity depends strongly on wavelength. To better highlight this, we compare three different wavebands in Figure~\ref{fig:optical_comparison}, where $500\,{\rm nm}$ roughly corresponds to $g$-band and $650\,{\rm nm}$ roughly corresponds to $r$-band. {This shows that longer wavelengths generally show less variation. This is consistent with the ensemble of light curves presented in \citep{Mummery2024}, which show that the higher cadence {\it g}- and {\it r}-band light curves generally show variations by a factor of $\sim3$ rather than being simply flat. This roughly matches the variations we see in Figure~\ref{fig:sed_multi}. \citet{Nicholl2024} also present late time {\it r}-band observations of AT2019qiz. The variations appear somewhat smaller, maybe a factor of $\sim2$, although there are large error bars due to subtracting the host galaxy from the TDE light. To facilitate better comparisons at these longer wavelengths, it will be useful to build one-dimensional models rather than the one-zone models we use here to better follow material at the largest radii where most of this light originates from.}

\subsection{Radio Flares}

As described in Section~\ref{sec:disksolutions}, the super-Eddington phases of the evolution typically ejecta $M_{\rm ej}\sim10^{-3}-10^{-1}\,M_\odot$. The material will have a variety of velocities, roughly scaling with the escape velocity from each radius in the disk. In the outer parts of the disk, $v_{\rm min}\sim0.03c$, but near the ISCO it can be $v_{\rm max}\sim0.3c$ or somewhat higher. {Using the outflow rate given in Equation~(\ref{eq:mdotout}), this results in a density profile,
\be
    \rho_{\rm out}(v,t) = \frac{p\dot{M}\Delta t}{2\pi (vt)^3}
    \lp \frac{v_{\rm min}}{v}\rp^{2p},
\ee
where $\Delta t$ is the timescale out of the outburst. Assuming a roughly spherical outflow, the optical depth of this material is
\be
    \tau(t) &=& \int_{v_{\rm min}t}^{v_{\rm max}t}\kappa \rho(v,t)d(vt)
    \nonumber
    \\
    &\approx&\frac{ p}{4\pi(p+1)}\frac{\kappa\dot{M}\Delta t}{(v_{\rm min}t)^2}, 
\ee
where we assume $v_{\rm min}\ll v_{\rm max}$. Setting $\tau\approx1$, we can solve for the time when the material becomes optically thin, which is only $\sim0.5\,{\rm days}$. Thus this material could reprocess the disk emission, and contribute to the optical/UV luminosity, but such a phase may be difficult to catch because it is so short.}

Another possibility is that this ejected material generates radio emission. As the ejecta continues to expand, it will interact with surrounding material with density $n$, sweeping up a mass comparable to its own on at a Sedov-Taylor radius of
\be
    r_{\rm TS} \approx \lp \frac{3M_{\rm flare}}{4\pi n m_p}\rp^{1/3},
\ee
where $m_p$ is the proton mass. A typical radius would be $r_{\rm TS}\sim10^{17}-10^{18}\,{\rm cm}$, depending on the value of $n$ which may be higher in the cicumnuclear environment. Thus we would expect radio synchrotron emission on a timescale of $\sim{\rm yrs}$. This is analogous to the radio flare model for neutron star mergers by \citet{Nakar2011}, and we would expect similar luminosities of \mbox{$\sim10^{38}\,{\rm erg\,s^{-1}}$} at around $\sim{\rm GHz}$ frequencies. Such a model may have difficulty though producing flares that fall off more quickly \citep[e.g., see some of the radio light curves in][]{Alexander2020}. This could instead require collisions between successive ejections when the fast outer material of a later shell catches the slower inner material of an earlier shell. We save a detailed calculation of this process for future work.

Roughly speaking though, when \citet{Cendes2024} fit synchrotron models to real events, they find similar parameters to what we find here. Furthermore, tracing the velocities of the radio emitting regions back in time requires material to be ejected after the main TDE on timescales similar to when we see disk instabilties. This all suggests that the super-Eddington flares our disk models exhibit could naturally explain the observed radio flares, which we will explore in future work. Since the amount of mass in the ejected depends on $M_{\rm BH}$, modeling the flares could provide a complementary way to constrain the BHs in TDEs.

\section{Discussion and Conclusions}
\label{Sec:conclusions}

We explored the evolution of TDE accretion disks on long timescales after the initial stellar destruction. Using a semi-analytic one-zone model, we showed that these disks naturally go into a flaring state due to a thermal instability and continued feeding by fallback accretion. The high super-Eddington state can last for a couple of days, ejecting $\sim10^{-3}-10^{-1}\,M_\odot$ of material at speeds of $\sim0.03-0.3c$, potentially producing radio flares that can occur up to a decade after the main TDE \citep[e.g.,][]{Cendes2024}. This is interspersed with low states where the disk luminosity rises for months to years as it grows in mass. This low state matches the late-time UV luminosities seen years after some TDEs \citep[e.g.,][]{vanVelzen19}. {The UV luminosity can vary by over an order of magnitude in the low state, so we may only be observing events at times when the disks are especially massive and bright. This is difficult to assess because current UV observations are fairly sparse. We find that the luminosity in the optical bands varies less, which is similar to the factor of $\sim3$ variations seen in the late time {\it r}- and {\it g}-band observations \citep{Mummery2024}.}

This work suggests there may be interesting connections between the disk state and the occurrence of radio flares. Continued monitoring in the optical/UV to characterize the disk and in the radio to follow the flares could be used to better constrain the time dependent accretion state \citep[e.g.,][]{Goodwin2024}. We demonstrate how the rate of flare events are related to the properties of the TDE, so in the future this may be used as another way to measure the BH mass. Although our preliminary work in Section~\ref{sec:fallbackheating} on the impact of fallback heating shows this could prevent flaring at early times, so this should be explored in more detail in future studies.

{Our $M_{\rm BH}=10^6\,M_\odot$ disk model transitions from the high state to the low state for the first time at $\sim100\,{\rm days}$, depending on the details of the parameters chosen (e.g., see Figure~\ref{fig:time_evol_short}). In contrast, the TDE AT2018fyk shows a sudden drop in X-ray emission at $\sim320$ and $\sim500\,{\rm days}$ \citep{Wevers2019,Wevers2023} while AT2021ehb shows a similar drop at $\sim320\,{\rm days}$ \citep{Yao2022}. This could potentially be matched by more massive BHs. Alternatively, \citet{Lu2022} speculates that these longer timescales could be easier to explain if the disk experiences an outside-in transition to a low accretion state. This process is estimated to roughly take $\sim1\,{\rm yr}$ (set by when the inner disk reaches the Eddington limit), which is in better agreement with AT2018fyk and AT2021ehb. Addressing whether this occurs would require a one-dimensional treatment rather than the one-zone model we employ here. While this is outside of the scope of this work, such a calculation could be fruitful for better understanding these observations. Another complication that we do not include is the impact of magnetic fields, which can help to stabilize the TDE disks \citep[e.g.][]{Alush2025}. Higher cadence observations, especially the the UV with missions like ULTRASAT \citep{ultrasat}, would be helpful for constraining whether such physical effects are needed.}

Our models also provide the disk properties that are present for QPEs in the star-disk collision picture. As the disk grows in the low state over many months to years, it would also presumably change the properties of the QPEs as well (duration, energy, etc). This could provide another probe of the disk physics. There could also be an additional mass source for the disk from material ablated from the QPE-producing star, and in fact \citet{Linial24} argue that heating from the star-disk collision could help stabilize the high and low states we explore here (similar to the fallback heating we explore). This suggests that the radio flaring and optical/UV disk emission could be impacted by QPEs, providing yet another reason to conduct long term multiwavelength monitoring of TDEs.

\acknowledgments

We thank Xiaoshan Huang, Daichi Tsuna, and Wenbin Lu for helpful discussions and feedback. {We also thank the referee for useful comments and suggestions.}

\begin{appendix}

\section{Solving for the Disk Evolution}
\label{app:solving}

To follow the time evolution of the disk, Equations~(\ref{eq:mass}) and (\ref{eq:angularmomentum}) are solved forward in time using explicit time-stepping. The accretion rate can vary by orders of magnitude during its evolution, so the time steps must be similarly flexible. This is done by setting the $i$th time step as
\be
    \Delta t_i
    = \epsilon
    \frac{M_d(t_{i-1})}{|\dot{M}_{\rm fb}(t_{i-1})
        - \dot{M}(t_{i-1})|},
    \label{eq:timestep}
\ee
where $\epsilon$ is a small number that adjusts the size of the time steps and all the parameters on the righthand side of Equation~(\ref{eq:timestep}) are evaluated at the previous time step.  We find good convergence for $\epsilon<10^{-3}$.

The main challenge when finding the evolution is solving for the accretion rate $\dot{M}$. This is done using the following strategy.
\begin{itemize}
    \item The surface density is solved for using the $M_d$ and $R_d$ from the previous timestep and $\Sigma=M_d/(\pi R_d^2)$.
    \item The temperature $T$ is then found by numerically solving the energy balance Equation~(\ref{eq:energy}). This is done using bisection around a temperature interval set from the temperature found in the previous time step, $T(t_{i-1})$, since for some values of $\Sigma$ there may be multiple solutions for $T$. 
    \item An important step when solving the energy equation is that for a given $\Sigma$ and guess for $T$, the isothermal sound speed can be solved with a quadratic equation $c_s^2-aT^4/(3\Sigma\Omega)c_s-k_{\rm B}T/(\mu m_p)=0$.
    \item Once $\Sigma$ and $T$ are found, we set $\dot{M}=3\pi\alpha c_s^2\Sigma/\Omega$.
\end{itemize}
At this point, we solve for $\Delta t_i$ using Equation~(\ref{eq:timestep}), and then we evolve forward one time step using
\be
    M_d(t_i) = M_d(t_{i-1}) + \dot{M}_{\rm fb}(t_{i-1})\Delta t_i - \dot{M}(t_{i-1})\Delta t_i,
\ee
and
\be
    J_d(t_i) = J_d(t_{i-1})
    + j_{\rm fb}\dot{M}_{\rm fb}(t_{i-1})\Delta t_i
    - C(GM_{\rm BH}R_d)^{1/2}\dot{M}(t_{i-1})\Delta t_i.
\ee
The new radius is then set by using $R_d=(J_d/M_d)^2/(GM_{\rm BH})$. We then go back to the first bullet point above to start working on the next time step. {Implicit in our approach is that the thermal time of the disk is always much shorter than the viscous time so that energy balance is always maintained. This prevents us from being forced to explicitly time evolve the internal energy of the disk. Our comparisons in Section~\ref{sec:comparison}, demonstrate that this assumption results in solutions that match previous work.}

\end{appendix}

\end{document}